\documentclass[prd,nofootinbib,showpacs,twocolumn,preprintnumbers,eqsecnum]{revtex4-1}
\usepackage{color}
\usepackage{bm}
\usepackage{graphicx}
\usepackage{amsmath}
\usepackage{amssymb}
\usepackage{enumitem}
\usepackage{hyperref}
\usepackage{subfigure}
\usepackage{array}
\usepackage[english]{babel}
\usepackage{tensor}
\usepackage{comment}
\usepackage[normalem]{ulem}
\usepackage[dvipsnames]{xcolor}

\allowdisplaybreaks

\def\be{\begin{equation}}
\def\ee{\end{equation}}
\def\ba{\begin{eqnarray}}
\def\ea{\end{eqnarray}}
\def\l{\left}
\def\r{\right}
\def\f{\frac}

\usepackage{tikz}

\def\nn{\nonumber}

\def\rd{{\rm d}}

\begin{document}

\title{{\bf Cosmological constraints and phenomenology of a beyond-Horndeski model} }

\author{Simone Peirone$^{1}$, Giampaolo Benevento$^{2,3,4}$, 
Noemi Frusciante$^{5}$, Shinji Tsujikawa$^{6}$\\}
\affiliation{ 
\smallskip
$^{1}$Institute Lorentz, Leiden University, PO Box 9506, Leiden 2300 RA, The Netherlands
\smallskip\\
$^{2}$Dipartimento di Fisica e Astronomia ``G. Galilei'', Universit\`a degli Studi di Padova, 
via Marzolo 8, I-35131, Padova, Italy
\smallskip\\
$^{3}$INFN, Sezione di Padova, via Marzolo 8, I-35131, Padova, Italy
\smallskip\\
$^{4}$Kavli Institute for Cosmological Physics, Department of Astronomy \& Astrophysics,
Enrico Fermi Institute, The University of Chicago, Chicago, IL 60637, USA
\smallskip\\
$^{5}$Instituto de Astrof\'isica e Ci\^encias do Espa\c{c}o, Faculdade de Ci\^encias da Universidade de Lisboa,  \\
 Edificio C8, Campo Grande, P-1749016, Lisboa, Portugal 
\smallskip \\
$^{6}$Department of Physics, Faculty of Science, Tokyo University of Science,
1-3, Kagurazaka, Shinjuku-ku, Tokyo 162-8601, Japan  }

\begin{abstract}
We study observational constraints on a  specific dark energy model in the framework 
of Gleyzes-Langlois-Piazza-Vernizzi theories, which extends the Galileon ghost 
condensate (GGC) to the domain of beyond Horndeski theories.
 In this model, we show that the Planck cosmic microwave background (CMB) data, combined 
with datasets of baryon acoustic oscillations, supernovae type Ia, and redshift-space 
distortions, give the tight upper bound 
$|\alpha_{\rm H}^{(0)}| \le {\cal O}(10^{-6})$ on today's 
beyond-Horndeski (BH) parameter $\alpha_{\rm H}$. 
This is mostly attributed to the shift of CMB acoustic peaks induced by the 
early-time changes of cosmological background and perturbations 
arising from the dominance of $\alpha_{\rm H}$ in the dark energy density. 
In comparison to the $\Lambda$-cold-dark-matter ($\Lambda$CDM) model, 
 our BH model suppresses the large-scale integrated-Sachs-Wolfe (ISW) tail of 
CMB temperature anisotropies due to the existence of cubic Galileons, 
and it modifies the small-scale CMB 
power spectrum because of the different background evolution.
We find that the BH model considered fits the data better than $\Lambda$CDM 
according to the $\chi^2$ statistics, yet the deviance information criterion (DIC) 
slightly favors the latter. Given the fact that  our BH model with 
$\alpha_{\rm H}=0$ (i.e.,  the GGC model) is favored over 
$\Lambda$CDM even by the DIC, there are no particular 
signatures for the departure from Horndeski theories 
in current observations.
\end{abstract}

\date{\today}

\maketitle

\section{Introduction}\label{Intro}

Despite the tremendous progress of observational cosmology over the past 
two decades, the origin of today's acceleration of the Universe has not been 
identified yet. The standard concordance scenario is 
the $\Lambda$CDM model, in which 
the cosmological constant $\Lambda$ 
is the source for cosmic acceleration. 
In addition to the difficulty of naturally explaining the origin of 
$\Lambda$ from the vacuum energy \cite{Weinberg,Martin,Tony}, 
it is known that there are tensions
between some datasets in the estimations of  today's 
value of the Hubble constant 
$H_0=100\,h$ km\,sec$^{-1}$\,Mpc$^{-1}$ \cite{Riess:2011yx,Delubac:2014aqe,Riess:2016jrr,Abbott:2017wau,Abbott:2017smn} and 
the amplitude $\sigma_8$ of the matter power spectrum on the scale of 
$8\,h^{-1}$~Mpc \cite{deJong:2015wca,Kuijken:2015vca,Hildebrandt:2016iqg,Conti:2016gav,Douspis:2018xlj}. 
Such observational tensions along with the theoretical shortcoming of $\Lambda$CDM reinforce the idea to look for alternative models of 
dark energy \cite{dreview1,dreview2,dreview3,dreview4,dreview5,dreview6,Kase:2018aps}.

Many extensions to the standard cosmological scenario include large-distance modifications 
of gravity due to an extra scalar degree of freedom (DOF), thus they are 
dubbed scalar-tensor theories \cite{Fujii:2003pa}.
Among those, the Horndeski class of theories \cite{Horndeski} is the most 
general scheme with second-order equations 
of motion \cite{Def11,KYY,Char11}. 
The latter feature ensures the absence of Ostrogradski 
instabilities, related to the existence of higher-order time derivatives.

It is possible to construct healthy theories beyond Horndeski gravity 
free from Ostrogradski instabilities.
In Gleyzes-Langlois-Piazza-Vernizzi (GLPV) theories \cite{GLPV}, 
for example, there are two extra Lagrangians beyond the Horndeski domain
without increasing the extra propagating DOFs \cite{Hami1,Hami2}.
GLPV theories have several peculiar properties: the propagation speeds of 
matter and the scalar field are mixed  \cite{Gergely:2014rna,Kase:2014yya,DeFelice:2015isa,DeFelice:2016ucp}, a partial breaking of the Vainshtein mechanism 
occurs inside astrophysical bodies~\cite{Kobayashi:2014ida,Koyama:2015oma,Sakstein:2015zoa,Sakstein:2015aac,Sakstein:2016ggl,Babichev:2016jom}, and 
a conical singularity can arise at the center of a spherically symmetric 
and static body \cite{DeFelice:2015sya,Kase:2015gxi}. 
We note that there exist also extensions of Horndeski theories 
containing higher-order spatial 
derivatives \cite{Gao:2014fra,Kase:2014cwa,Frusciante:2015maa} 
(encompassing Horava gravity \cite{Horava}) and degenerate 
higher-order scalar-tensor theories with one scalar 
propagating DOF \cite{Langlois1,Langlois2,CKT,Motohashi:2016ftl}.

The detection of the gravitational wave (GW) signal 
GW170817 \cite{GW170817} accompanied by the gamma-ray burst event GRB170817A \cite{Goldstein} shows that  
the speed of GWs $c_t$ is constrained to be
in the range $-3\times 10^{-15}\leq c_t-1\leq 7\times 10^{-16}$ \cite{Abbott} 
at the redshift $z \le 0.009$, where we use the unit in which the speed of 
light $c$ is equivalent to 1. The Horndeski Lagrangian, which gives 
the exact value $c_t=1$ without the tuning among functions, 
is of the form $L=G_4(\phi)R+G_2(\phi,X)+G_3(\phi,X) \square \phi$, 
where 
$G_4$ is a function of the scalar field $\phi$, $R$ is the Ricci scalar, 
and $G_{2,3}$ depend on both $\phi$ and 
$X=\partial^{\mu} \phi \partial_{\mu} \phi$ \cite{GWcon1,GWcon2,GWcon3,GWcon4,GWcon5,GWcon6}. 
There are also dark energy models in which the GW speed
consistent with the above observational bound of $c_t$ 
can be realized \cite{Battye:2018ssx,Amendola:2018ltt,Copeland:2018yuh}.
In GLPV theories with the $X$ dependence in $G_4$, it is also possible to 
realize $c_t=1$ by the existence of an additional quartic Lagrangian 
beyond the Horndeski domain \cite{Kase:2018iwp}.

In addition to the bound on $c_t$, the absence of the decay of GWs into dark energy  
at LIGO/Virgo frequencies ($f \sim 100$~Hz) may imply that the parameter 
$\alpha_{\rm H}$ characterizing the deviation from Horndeski theories is 
constrained to be very tiny
for the scalar sound speed $c_s$ different from 1, typically of order 
$|\alpha_{\rm H}| \lesssim 10^{-10}$ today \cite{Creminelli:2018xsv}. 
If we literally use this bound, there is little room left for 
dark energy models in beyond-Horndeski theories \cite{Frusciante:2018tvu,Kobayashi:2019hrl}. 
 If $c_s$ is equivalent to 1, the decay of GWs into dark energy is forbidden. 
However, it was 
argued in Ref.~\cite{Creminelli:2018xsv} that power-law divergent terms would appear, 
leading to the conclusion that the operator accompanying $\alpha_{\rm H}$ 
must be suppressed as well \cite{Creminelli:2018xsv}.

We note that the LIGO/Virgo frequencies are close to those of the typical 
strong coupling scale or cut-off $\Lambda_c$ of dark energy models containing derivative 
field self-interactions \cite{deRham:2018red}. 
Around this cut-off scale, we cannot exclude the possibility that some ultra-violet (UV) 
effects come into play to recover the propagation and property of GWs similar to those 
in General Relativity (GR). If this kind of UV completion occurs around the frequency 
$f \sim 100$~Hz, the aforementioned bounds on $c_t$ and $\alpha_{\rm H}$ are not 
applied to the effective field theory of dark energy 
exploited to describe the cosmological 
dynamics much below the energy scale $\Lambda_c$.
Future space-based missions, such as LISA \cite{Audley:2017drz}, are sensitive to much
lower frequencies ($f \sim 10^{-3}$~Hz), so they will offer further valuable information 
on the properties of GWs with different frequencies.

In GLPV theories, there are constraints on the parameter $\alpha_{\rm H}$ arising from 
the modifications to gravitational interactions inside astrophysical objects. 
 For example, the consistency of the minimum mass for hydrogen burning in stars with the red dwarf 
of lowest mass shows that $|\alpha_{\rm H}|$ is at most of 
order 0.1 \cite{Sakstein:2015zoa,Sakstein:2015aac,Jain:2015edg,Saltas:2018mxc}.
By using X-ray and lensing profiles of galaxy clusters, similar bounds on 
$\alpha_{\rm H}$ were obtained in Ref.~\cite{Sakstein:2016ggl}.
From the orbital period of the Hulse-Taylor binary pulsar PSR B1913+1, 
the upper bound of $|\alpha_{\rm H}|$ is of order $10^{-3}$ \cite{Dima:2017pwp}. 
 Cosmological constraints on $\alpha_{\rm H}$ were derived by using particular 
parametric forms of dimensionless quantities appearing in the effective field theory of 
dark energy to describe their evolution. 
In this case, the constraints from CMB and large scale structure data on 
$|\alpha_{\rm H}|$ 
are of order $\mathcal{O}(1)$ \cite{Traykova:2019oyx}.

In this paper, we place observational bounds on the beyond-Horndeski (BH) 
dark energy model proposed in Ref.~\cite{Kase:2018iwp} and study how 
the parameter $\alpha_{\rm H}$ 
is constrained from the cosmological datasets of CMB temperature 
anisotropies, baryon acoustic oscillations (BAO), supernovae type Ia (SN Ia), 
and redshift-space distortions (RSDs). 
In the limit $\alpha_{\rm H} \to 0$, the model 
reduces to the Galileon ghost condensate (GGC) 
in Horndeski theories. 
The recent analysis of Ref.~\cite{PBFT} reveals that the 
GGC model is observationally favored over $\Lambda$CDM according to several information criteria. 
We will investigate whether or not this property persists for the BH 
dark energy model ($\alpha_{\rm H} \neq 0$) of Ref.~\cite{Kase:2018iwp}. 
For the likelihood analysis, we will use the publicly available 
Effective-Field-Theory for CAMB (EFTCAMB) 
code\footnote{Web page: \url{http://www.eftcamb.org}} \cite{Hu:2013twa,Raveri:2014cka}.
 In our investigation the gravitational theory is completely determined by a covariant action, while the analysis in Ref. \cite{Traykova:2019oyx} follows a parameterized approach to GLPV theories.  
In this respect, the two cosmological models considered are completely different 
and the constraint on $\alpha_{\rm H}$ obtained in this paper cannot be 
straightforwardly compared to the results in Ref. \cite{Traykova:2019oyx}.

The paper is organized as follows. 
In Sec~\ref{Sec:theory}, we briefly review the basics of the BH dark energy 
model introduced in Ref.~\cite{Kase:2018iwp}. 
In Sec.~\ref{Sec:methodology}, we show how this model can be 
implemented in the EFT formulation 
and derive the background equations of motion together with 
theoretically consistent conditions. 
In Sec.~\ref{Sec:perturbations}, we discuss the evolution of cosmological 
perturbations in the presence of matter perfect fluids 
and investigate the impact of our model 
on observable quantities.
In Sec.~\ref{Sec:constraints}, we present the Monte-Carlo-Markov-Chain (MCMC) 
constraints on model parameters and compute several information criteria to discuss 
whether the BH model is favored over the $\Lambda$CDM model.
Finally, we conclude in Sec.~\ref{Sec:conclusion}.

\section{Dark energy model in GLPV theories}\label{Sec:theory}

The dark energy model proposed in Ref.~\cite{Kase:2018iwp} belongs to 
the quartic-order GLPV theories given by the action 
\be
\mathcal{S}=\int{}\rd^4x\sqrt{-g}\sum^4_{i=2}L_i
+{\cal S}_M[g_{\mu\nu},\chi_M]\,,
\label{action}
\ee
where $g$ is the determinant of metric tensor $g_{\mu\nu}$, 
${\cal S}_M$ is the matter action for all matter fields $\chi_M$,
and the Lagrangians $L_{2,3,4}$ are defined by
\ba
L_2 &=& G_2(\phi,X)\,,\nn\\
L_3 &=& G_3(\phi,X)\Box\phi\,,\nn\\
L_4 &=& G_{4}(\phi,X) R-2G_{4,X}(\phi,X)
\left[ (\square \phi )^{2}-\phi^{\mu\nu} 
\phi_{\mu\nu} \right] \nn\\
&&+ F_4(\phi,X)\epsilon^{\mu\nu\rho}_{\phantom{\mu\nu\rho}\sigma}\epsilon^{\mu'\nu'\rho'\sigma}\phi_{\mu'}\phi_\mu\phi_{\nu\nu'}\phi_{\rho\rho'}\,,
\ea
where  $G_{2,3,4}$ and $F_{4}$ are functions of the scalar field $\phi$ and 
$X=\nabla^\mu\phi \nabla_\mu\phi$, $R$ is the Ricci scalar, and 
$\epsilon^{\mu\nu\rho\sigma}$ is the totally antisymmetric Levi-Civita tensor 
satisfying the normalization 
$\epsilon^{\mu\nu\rho\sigma}\epsilon_{\mu\nu\rho\sigma}=+4!$.
We also define $G_{i,X}\equiv \partial G_i/\partial X$ and use 
the notations $\phi_\mu=\nabla_\mu\phi$ and $\phi_{\mu\nu}=\nabla_\nu\nabla_\mu\phi$ 
for the covariant derivative operator $\nabla_\mu$. 
We assume that the matter fields $\chi_M$ are minimally coupled to gravity.

The last term containing $F_4(\phi,X)$ in $L_4$ arises beyond the domain of 
Horndeski theories \cite{GLPV}. 
The deviation from Horndeski theories can be quantified by 
the parameter
\be
\label{aH}
\alpha_{\rm H}=-\frac{X^2 F_4}{G_4-2XG_{4,X}+X^2 F_4}\,,
\ee
which does not vanish for $F_4 \neq 0$.
The line element containing intrinsic tensor perturbations $h_{ij}$ 
on the flat Friedmann-Lema\^{i}tre-Robertson-Walker (FLRW) space-time
is given by 
\be
{\rm d} s^2=-{\rm d}t^2+a^2(t) \left( \delta_{ij}
+h_{ij} \right) {\rm d}x^i {\rm d}x^j\,,
\ee
where $a(t)$ is the time-dependent scale factor, and $h_{ij}$ 
satisfies the transverse and traceless conditions 
($\nabla^j h_{ij}=0$ and ${h_i}^i=0$).
The propagation speed squared of 
tensor perturbations is \cite{GLPV,Gergely:2014rna,Kase:2014yya}
\be
c_t^2=\frac{G_4}{G_4-2XG_{4,X}+X^2 F_4}\,.
\ee
In quartic-order Horndeski theories ($F_4=0$), the $X$ dependence in $G_4$ 
leads to the difference of $c_t^2$ from 1. 
In GLPV theories, it is possible to realize $c_t^2=1$ 
for the function 
\be
F_4=\frac{2G_{4,X}}{X}\,,
\label{F4con}
\ee
under which $\alpha_{\rm H}=-2XG_{4,X}/G_4$.

In this paper, we will study observational constraints on the model proposed 
in Ref.~\cite{Kase:2018iwp}. 
This is characterized by the following functions
\ba
\label{model}
G_2&=&a_1X+a_2X^2\,,\qquad G_3=3a_3X\,, \nn \\ 
G_4&=&\frac{m_0^2}{2}-a_4X^2\,,\qquad F_4=-4a_4\,, 
\ea
where $m_0$ and $a_{1,2,3,4}$ are constants. 
This beyond-Horndeski model, hereafter BH, satisfies 
the condition (\ref{F4con}), and hence $c_t^2=1$.
When $a_4=0$, BH recovers the GGC model 
studied recently in Ref.~\cite{PBFT}. 
Taking the limits $a_2 \to 0$ and $a_3 \to 0$, GGC recovers 
the cubic covariant Galileon \cite{Nicolis,Galileons} and 
ghost condensate \cite{Arkani}, respectively.

The BH model allows for the existence of self-accelerating de Sitter 
solutions finally approaching constant values of $X$. 
Before approaching the de Sitter attractor, the dark energy equation of state $w_{\rm DE}$ 
can exhibit a phantom behavior (i.e., $w_{\rm DE}<-1$) 
without the appearance of ghosts \cite{Kase:2018iwp}. 
The cubic covariant Galileon gives rise to the tracker solution with $w_{\rm DE}=-2$ 
in the matter era \cite{DT10}, but this evolution is incompatible 
with the joint data analysis of CMB, BAO, and SN Ia \cite{NDT10}. 
On the other hand, in both BH and GGC, the $a_2 X^2$ term works to prevent for 
approaching the tracker, so that $-2<w_{\rm DE}<-1$ in the matter era.
This behavior of $w_{\rm DE}$ is consistent with the recent 
observational datasets of CMB, BAO, and SN Ia \cite{PBFT}.

The BH model leads to the evolution of cosmological perturbations different from 
that in GR. The late-time modification to the cosmic growth rate arises mostly from
the cubic Galileon term $3a_3X \square \phi$ \cite{Kase:2018iwp, Bartolo}.
In GGC, the combined effect of $3a_3X \square \phi$ and $a_2 X^2$ can suppress 
the power spectrum of large-scale CMB temperature anisotropies, so that the model 
shows a better compatibility with the Planck data with respect to the 
$\Lambda$CDM \cite{PBFT}. 
It remains to be seen whether the similar property also holds for the BH model 
with $a_4 \neq 0$, 
which we will address in this paper.

\section{Methodology}
\label{Sec:methodology}

In this section, we discuss the evolution of the background and linear scalar perturbations in the BH model. We make use of the EFTCAMB/EFTCosmoMC codes \cite{Hu:2013twa,Raveri:2014cka}, 
in which the EFT of dark energy and modified 
gravity \cite{Gubitosi:2012hu,Bloomfield:2013efa,Bloomfield:2012ff,Gleyzes:2013ooa,Gleyzes:2014rba} 
is implemented into CAMB/CosmoMC \cite{Lewis:1999bs,Lewis:2002ah}. 
The EFT framework enables one to deal with any dark energy 
and modified gravity model with one scalar propagating DOF $\phi$ in a unified and model-independent manner.

The EFT of dark energy is based on the 3+1 Arnowitt-Deser-Misner (ADM) 
decomposition of spacetime \cite{ADM} given by the line element 
\be
\rd s^2=-N^2 \rd t^2+h_{ij} \left( \rd x^i +N^i \rd t \right) 
\left( \rd x^j +N^j \rd t \right)\,,
\ee
where $N$ is the lapse, $N^i$ is the shift vector, and $h_{ij}$ 
is the three-dimensional metric. 
A unit vector orthogonal to the constant time hyper-surface $\Sigma_t$ 
is given by $n_{\mu}=N \nabla_{\mu}t=(N, 0,0,0)$.
The extrinsic curvature is defined by $K_{ij}=
h^{k}_{i} \nabla_{k} n_{j}$. 
The internal geometry of $\Sigma_t$ is quantified by the three-dimensional 
Ricci tensor ${\cal R}_{ij}={}^{(3)}R_{ij}$ associated with the metric $h_{ij}$.

On the flat FLRW background, we consider the line element containing 
three scalar metric perturbations $\delta N$, $\psi$, and $\zeta$, as 
\ba
\rd s^2 &=&-(1+2\delta N) \rd t^2+2\partial_{i} \psi 
\rd t \rd x^i \nonumber \\
& &
+a^2(t)(1+2\zeta)  \delta_{ij}
{\rm d}x^i {\rm d}x^j\,,
\label{permet}
\ea
where $\partial_{i} \equiv \partial/\partial x^i$.
We also choose the the unitary gauge in which the perturbation $\delta \phi$ of 
the scalar field $\phi$ vanishes.  Then, the perturbations of extrinsic and intrinsic 
curvatures are expressed as \cite{Gleyzes:2013ooa,Gergely:2014rna,Kase:2014cwa,Gleyzes:2014rba}
\ba
\delta K_{ij} &=& a^2 \left( H \delta N-2H \zeta-\dot{\zeta} 
\right)\delta_{ij}+\partial_i \partial_j \psi\,,\\
\delta {\cal R}_{ij}
&=&-\delta_{ij} \partial^2 \zeta-\partial_i \partial_j \zeta\,,
\ea
where $\partial^2 \equiv \delta^{kl} \partial_k \partial_l$, and 
$H=\dot{a}/a$ is the Hubble expansion rate, and 
a dot represents a derivative with respect to $t$.
The perturbations of traces $K \equiv {K^{i}}_{i}$ and ${\cal R} 
\equiv {{\cal R}^i}_i$ are denoted as 
$\delta K$ and $\delta {\cal R}$, respectively, with 
$\delta g^{00}=2\delta N$.

In the ADM language, the Lagrangian of GLPV theories 
depends on the scalar quantities $N$, $K$, 
$K_{ij}K^{ij}$, ${\cal R}$, $K_{ij}{\cal R}^{ij}$, 
and $t$ \cite{Gleyzes:2013ooa}. 
Expanding the corresponding action up to second order in 
scalar perturbations of those quantities, it follows that 
\begin{align}
\label{eftaction}
\mathcal{S} =& \int \rd^4x \sqrt{-g}\,m_0^2 \bigg\{ \frac{1}{2} \left[1+\Omega(a)\right] R 
+ \f{\Lambda(a)}{m_0^2} - \f{c(a)}{m_0^2} \delta g^{00}  \nonumber\\
&+ H_0^2 \frac{\gamma_1 (a)}{2} \left( \delta g^{00} \right)^2
- H_0 \frac{\gamma_2(a)}{2} \, \delta g^{00}\,\delta K\nn\\
&-H_0^2 \frac{\gamma_3 (a)}{2} \left( \delta K \right)^2
- H_0^2 \frac{\gamma_4(a)}{2} \, \delta K^{i}_{j} \delta K^{j}_{i}\nn\\
&+\f{\gamma_5(a) }{2}\delta g^{00}\delta {\cal R}
\bigg\}+{\cal S}_M[g_{\mu \nu}, \chi_M]\,,
\end{align}
where $m_0$ is a constant having a dimension of mass, and
$\Omega$, $\Lambda$, $c$, $\gamma_{i}$ are called EFT functions 
that depend on the background scale factor $a$. 
The explicit relations between those EFT functions and the functions 
$G_{2,3,4}, F_4$ in the action (\ref{action}) are given in Ref.~\cite{Frusciante:2016xoj}.

The first three variables $\Omega$, $\Lambda$, $c$ determine  
both the background evolution and linear perturbations, 
whereas the functions $\gamma_{i}$ solely appear at the level of linear perturbations. 
For the matter action ${\cal S}_M$, we take dark matter and baryons 
(background density $\rho_m$ and vanishing pressure) and radiation 
(background density $\rho_r$ and pressure $P_r=\rho_r/3$) into account. 
Then, the background equations are expressed
as  \cite{Gubitosi:2012hu,Bloomfield:2012ff} 
\ba
&&3m_0^2H^2=\rho_{\rm DE}+\rho_m+\rho_r\,,\label{fried1}\\
&&-m_0^2 \left( 2\dot{H}+3H^2 \right)=
P_{\rm DE}+P_r \,,\label{fried2}
\ea
where 
\ba
\hspace{-0.5cm}
\rho_{\rm DE} &=& 2c-\Lambda-3m_0^2 H \left( \dot{\Omega}
+H \Omega \right) \,,\label{rho0}\\
\hspace{-0.5cm}
P_{\rm DE} &=& \Lambda+m_0^2 \left[ \ddot{\Omega}
+2H \dot{\Omega}+\Omega \left( 2\dot{H}+3H^2 \right) 
\right].\label{P0}
\ea
The density $\rho_{\rm DE}$ and pressure $P_{\rm DE}$ 
of dark energy obey the continuity equation 
\be
\label{scalarfield}
\dot{\rho}_{\rm DE}+3H \left( \rho_{\rm DE}
+P_{\rm DE} \right)=0\,.
\ee
In GLPV theories, there is the specific relation $\gamma_3=-\gamma_4$. 
If we restrict the theories to those satisfying $c_t^2=1$, 
it follows that  $\gamma_4=0$. 
Then, the model given by the functions (\ref{model}) 
corresponds to the coefficients
\be
\gamma_3=0\,,\qquad \gamma_4=0\,,
\ee
so that we are left with three functions $\gamma_1, \gamma_2, \gamma_5$ 
at the level of linear perturbations.

To study the cosmological evolution of our model in EFTCAMB, we first 
solve the background equations of motion and then map to the EFT functions 
according to the procedure given in 
Refs.~\cite{Gubitosi:2012hu,Bloomfield:2012ff,Bloomfield:2013efa,Gleyzes:2013ooa,Gleyzes:2014rba,Frusciante:2015maa,Frusciante:2016xoj}. 

\subsection{Background equations in the BH model}
\label{Sec:background}

For the model (\ref{model}), the background equations are given by 
Eqs.~(\ref{fried1}) and (\ref{fried2}), with 
\be
\Omega=-\frac{2a_4 \dot{\phi}^4}{m_0^2}\,,
\ee
and 
\ba
\hspace{-0.6cm}
\rho_{\rm DE}
&=&-a_1\dot{\phi}^2+3a_2\dot{\phi}^4+18a_3H\dot{\phi}^3+30a_4H^2\dot{\phi}^4\,,
\label{rhoDE} \\
\hspace{-0.6cm}
P_{\rm DE}
&=&-a_1\dot{\phi}^2+a_2\dot{\phi}^4-6a_3\dot{\phi}^2\ddot{\phi} \nn\\
\hspace{-0.6cm}
&&-2a_4\dot{\phi}^3\l[8H\ddot{\phi}+\dot{\phi}(2\dot{H}+3H^2)\r]\,.
\label{PDE}
\ea
The parameters $c$ and $\Lambda$ in Eqs.~(\ref{rho0}) and (\ref{P0}) 
can be expressed in terms of quantities on the right hand sides of 
Eqs.~(\ref{rhoDE}) and (\ref{PDE}). 
Following Ref.~\cite{Kase:2018iwp}, we define 
the dimensionless variables (density 
parameters):
\ba
&&x_1=-\f{a_1\dot{\phi}^2}{3m_0^2H^2}\,,
\qquad x_2=\f{a_2\dot{\phi}^4}{m_0^2H^2}\,,\nn\\
&&x_3=\f{6a_3\dot{\phi}^3}{m_0^2H}\,, 
\qquad x_4=\f{10a_4\dot{\phi}^4}{m_0^2}\,,
\label{dimensionless_functions}
\ea
and 
\be
\Omega_{\rm DE}=\frac{\rho_{\rm DE}}{3m_0^2H^2}\,,
\quad 
\Omega_m=\frac{\rho_{m}}{3m_0^2H^2}\,,
\quad 
\Omega_r=\frac{\rho_{r}}{3m_0^2H^2}\,.
\ee
{}From Eq.~(\ref{fried1}), we have 
\be
\Omega_{m}=1-\Omega_{\rm DE}-\Omega_{r}\,,
\label{constraint}
\ee
where the dark energy density parameter is 
given by 
\be 
\label{omega_constraint}
\Omega_{\rm DE}=x_1+x_2+x_3+x_4\,. 
\ee
In terms of $x_4$, the deviation parameter (\ref{aH}) from 
Horndeski theories is expressed as 
\be
\alpha_{\rm H}=\frac{4x_4}{5-x_4}\,,
\ee 
and hence $\alpha_{\rm H}$ is of the same order as $x_4$ 
for $|x_4| \le 1$.

The variables $x_{1,2,3,4}$ and $\Omega_r$ are known 
by solving the ordinary differential equations
\ba
&&x_1'=2x_1(\epsilon_\phi-h)\,,\label{system}\\
&&x_2'=2x_2(2\epsilon_\phi-h)\,,\\
&&x_3'=x_3(3\epsilon_\phi-h)\,,\\
&&x_4'=4x_4\epsilon_\phi\,,\\
&&\Omega_r'=-2\Omega_r(2+h)\,,
\label{system4}
\ea
where a prime denotes the derivative with respect to 
${\cal N}=\ln (a)$. On using Eqs.~(\ref{fried1}) 
and (\ref{fried2}), it follows that 
\ba
\epsilon_{\phi} 
&\equiv& 
\frac{\ddot{\phi}}{H \dot{\phi}} \nn\\
&=&-\frac{1}{q_s}\Bigg[20(3x_1+2x_2)-5x_3(3x_1+x_2+\Omega_r-3) \nn\\
&&- x_4(36x_1+16x_2+3x_3+8\Omega_r)\Bigg]\,, \nn
\label{epphi}\\
h 
&\equiv& \frac{\dot{H}}{H^2}   \nn\\
&=&
-\frac{1}{q_s}
[10(3x_1+x_2+\Omega_r+3)(x_1+2x_2) \nn \\
&& +10x_3(6x_1+3x_2+\Omega_r+3)+15x_3^2 \nn \\
& &+x_4(78x_1+32x_2+30x_3+12\Omega_r+36)
+12x_4^2]\,, \nn
\label{dh}
\ea
with 
\be
q_s \equiv 20(x_1+2x_2+x_3)
+4x_4 (6-x_1-2x_2+3x_3)
+5x_3^2+8x_4^2\,.
\label{qs}
\ee
For a given set of initial conditions $x_{1,2,3,4}$ and $\Omega_r$, 
we can solve Eqs.~(\ref{system})-(\ref{system4}) to determine 
the evolution of density parameters as well as $\phi$ and $H$.
Practically, we start to solve the above dynamical system at 
redshift $z_s=1.5 \times 10^5$ and iteratively scan over initial conditions leading to 
the viable cosmology satisfying the constraint (\ref{constraint}) today ($z=0$).
Additionally, evaluating Eq.~\eqref{omega_constraint} at present time, 
we can eliminate one model parameter, for example $x_2^{(0)}$, as 
$x_2^{(0)}=\Omega_{\rm DE}^{(0)}-x_1^{(0)}-x_3^{(0)}-x_4^{(0)}$, 
where ``(0)'' represents today's quantities.

\subsection{Mapping}\label{Sec:mapping}

To study the evolution of scalar perturbations and observational constraints on dark energy 
models in EFTCAMB, it is convenient to use the mapping 
between EFT functions and model parameters in BH. 
In Sec.~\ref{Sec:background}, we already discussed the mapping of the 
background quantities $\Omega, \Lambda$ and $c$. 
The functions $\gamma_{1,2,5}$, which are associated with scalar perturbations, 
are given by 
\ba
\gamma_1 &=& \frac{H^2}{ H_0^2}\l[\f{1}{20} 
\left(24 x_4- h x_4'+3 x_4'- x_4''\right)\r.
\nn\\
&&\l.\qquad 
+2 x_2
+\frac{1 }{12}\left\{ \left( h+9\right) x_3+ x_3'\right\} \r],\\
\gamma_2 &=& \frac{H}{H_0}\l[ \f{1}{5}\left(x_4'-8 x_4\right) - x_3\r], \\
\gamma_5 &=& \frac{2}{5} x_4.
\label{EFTmapping}
\ea
The expressions of these EFT functions allow us to draw already some insight about the contributions of each $x_i$ to the dynamics of linear perturbations. 
In general, the variable $\gamma_1$ cannot be well constrained by data being 
its contribution to the observables below the cosmic variance \cite{Frusciante:2018jzw}. 
The main modification to the evolution of perturbations compared to GR arises from 
$\gamma_{2}$ and $\gamma_5$, which are mostly affected by 
$x_3$ and $x_4$. The variables $x_1$ and $x_2$ contribute to the perturbation 
dynamics through the Hubble expansion rate $H$ in $\gamma_2$.

\subsection{Viability constraints}\label{Sec:theoconstraints}
 
There are theoretically consistent conditions under which the perturbations 
are not plagued by the appearance of ghosts and Laplacian instabilities 
in the small-scale limit.
For the BH model (\ref{model}), the conditions for the absence of ghosts 
in tensor and scalar sectors are given, respectively, by \cite{Kase:2018iwp}
\ba
Q_t &=& \frac{5-x_4}{10} m_0^2>0\,,\label{Qt} \\
Q_s &=& \frac{3(5-x_4)q_s}{25(x_3+2x_4-2)^2} m_0^2>0\,,
\label{Qs}
\ea
where $q_s$ is defined in Eq.~(\ref{qs}).
Then, we have the following constraints 
\be
x_4<5\,,\qquad q_s>0\,.
\ee
The BH model has the property $c_t^2=1$, so there is 
no Laplacian instability for tensor perturbations.  
We note that the reduced Planck mass $M_{\rm pl}$ is 
related to $m_0$ according to the relation 
$M_{\rm pl}^2=m_0^2 (1+\Omega_0)$ in the 
local environment with screened fifth 
forces \cite{Kobayashi:2014ida}, where 
$\Omega_0$ is today's value of $\Omega$. 
Then, the Newton gravitational constant $G_{\rm N}$ is 
given by 
\be
G_{\rm N}=\frac{1}{8\pi M_{\rm pl}^2}=
\frac{1}{8\pi m_0^2} \left( 1-\frac{x_4^{(0)}}{5} 
\right)^{-1}\,,
\label{GN}
\ee
which is positive under the absence of tensor ghosts.

For scalar perturbations, there are three propagation speed squares 
$c_s^2$, $\tilde{c}_r^2$, and $\tilde{c}_m^2$ associated with the 
scalar field $\phi$, radiation, and nonrelativistic matter, respectively.  
In Horndeski theories, they are not coupled to each other, so that the 
propagation speed squares of  radiation and nonrelativistic matter
are given, respectively, by $c_r^2=1/3$ and $c_m^2=+0$. 
In GLPV theories, they are generally mixed with 
each other, apart from $\tilde{c}_m^2$ 
(which has the value $\tilde{c}_m^2=+0$) \cite{GLPV,Gergely:2014rna,Kase:2014yya,DeFelice:2015isa,DeFelice:2016ucp}.
Then, the Laplacian instabilities of scalar perturbations can be avoided 
under the two conditions
\ba
&&c_s^2=\f{1}{2}\l(c_r^2+c_{\rm H}^2-\beta_{\rm H}-\gamma_{\rm H} 
\right)>0\,,\label{cs}\\
&&\tilde{c}_r^2=\f{1}{2}\l(c_r^2+c_{\rm H}^2-\beta_{\rm H}
+\gamma_{\rm H} \r)>0\,,
\label{crcon}
\ea
where
\ba
&&c_{\rm H}^2=\f{2}{Q_s}\l[\dot{\mathcal{M}}+H\mathcal{M}
-Q_t-\f{3\rho_m+4\rho_r}{12H^2(1+\alpha_{\rm B})^2}\r]\,,\nn\\
&&\beta_r=\f{4\alpha_{\rm H} \rho_r}{3Q_sH^2(1+\alpha_{\rm B})^2}\,,\quad 
\beta_m=\f{\alpha_{\rm H}\rho_m}{Q_s H^2(1+\alpha_{\rm B})^2}\,,\nn\\
&&\mathcal{M}=\f{Q_t(1+\alpha_{\rm H})}{H(1+\alpha_{\rm B})},\quad 
\beta_{\rm H}=\beta_r+\beta_m, \quad 
\alpha_{\rm B}=-\frac{5x_3+8x_4}{2(5-x_4)},\nn\\
&&\gamma_{\rm H}=\sqrt{(c_r^2-c_{\rm H}^2+\beta_{\rm H})^2
+2c_r^2\alpha_{\rm H} \beta_r}\,. 
\ea
When $|\alpha_{\rm H}| \ll 1$ we have $c_s^2 \simeq c_{\rm H}^2-\beta_{\rm H}$ 
and $\tilde{c}_r^2 \simeq c_r^2=1/3$, so the second stability condition 
(\ref{crcon}) is satisfied.

\begin{figure}
\includegraphics[width=.49\textwidth]{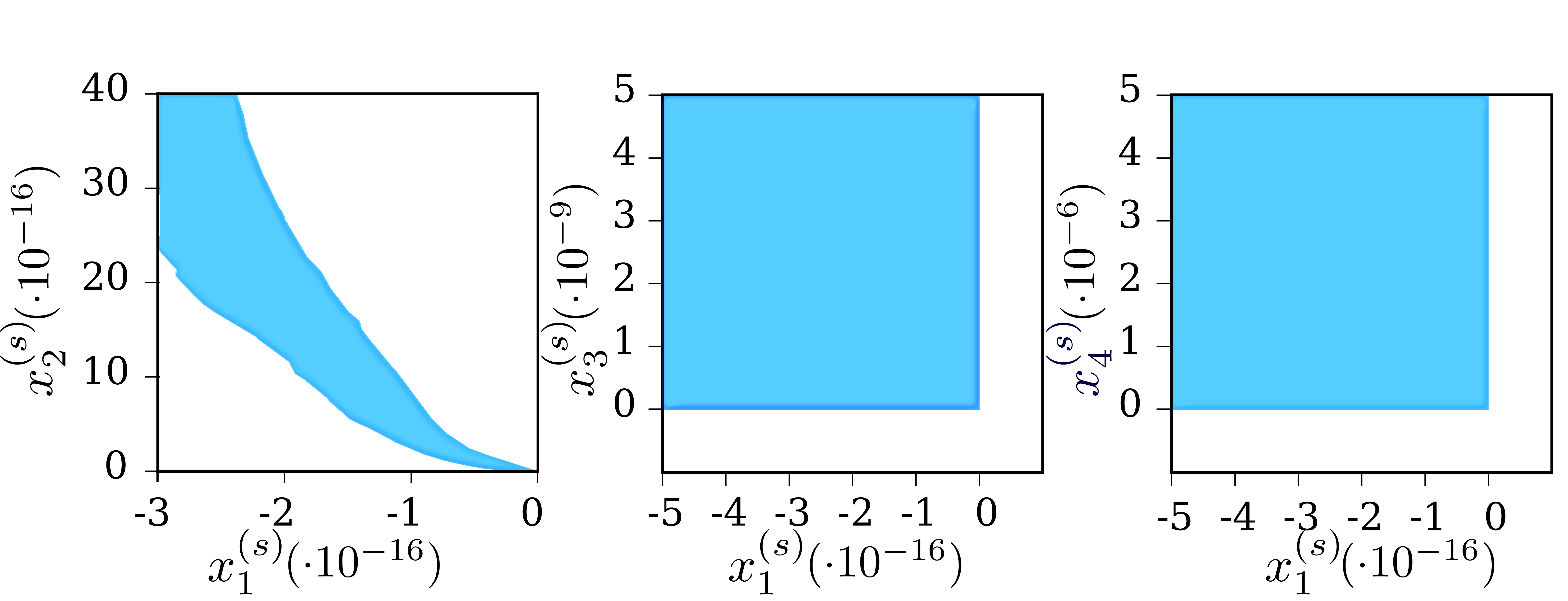}\\
\includegraphics[width=.49\textwidth]{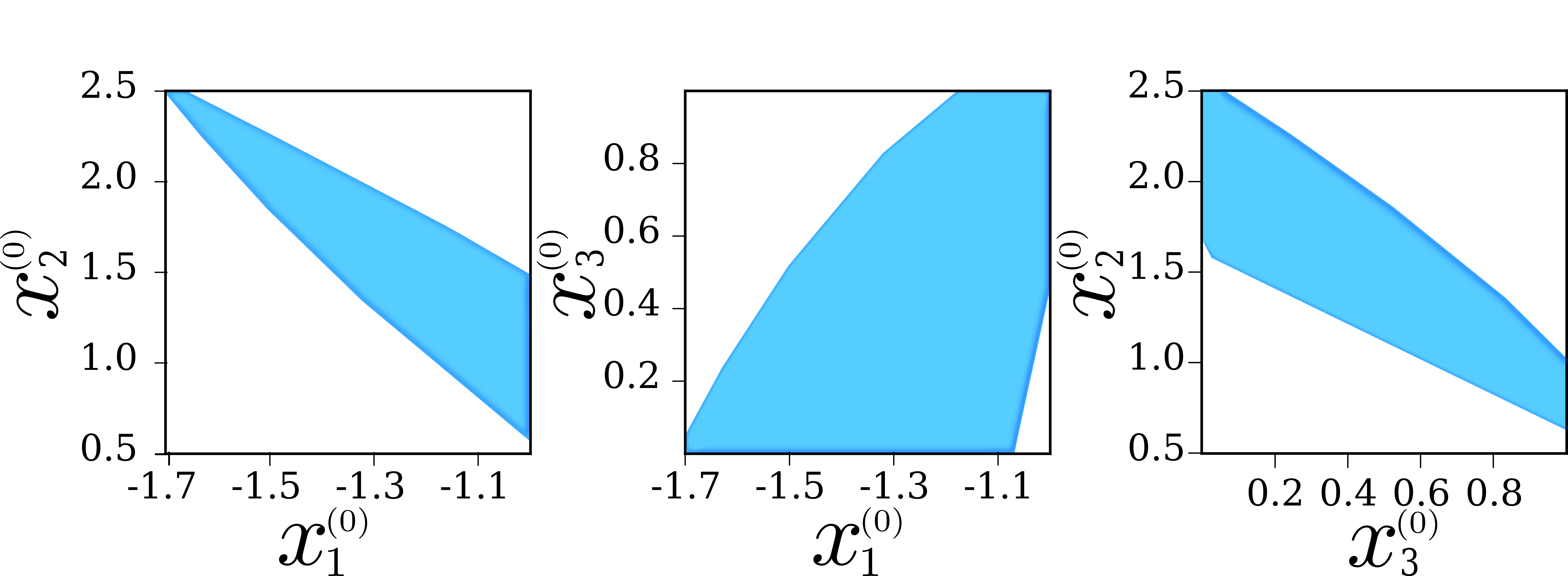}
\caption{\label{fig:stability} 
The viable parameter space (in blue) for the initial values 
$x_1^{(s)}$, $x_2^{(s)}$, $x_3^{(s)}$ and $x_4^{(s)}$ at the redshift 
$z_s=1.5 \times 10^{5}$ (top panel) and today's parameters $x_1^{(0)}$, $x_2^{(0)}$ and $x_3^{(0)}$ 
(bottom panel). In these parameter spaces, there are neither ghosts nor 
Laplacian instabilities. 
}
\end{figure}

There are also constraints on today's parameter $\alpha_{\rm H}^{(0)}$
(or equivalently, $x_4^{(0)}$) from massive astrophysical 
objects \cite{Sakstein:2015zoa,Sakstein:2015aac,Dima:2017pwp}. 
Among those constraints, the orbital period of Hulse-Taylor binary pulsar gives the 
tightest bound $-0.0031\leq x_4^{(0)}\leq 0.0094$ \cite{Dima:2017pwp,Kase:2018iwp}.
If we literally use the bound arising from 
the absence of the GW decay into dark energy  
at LIGO/Virgo frequencies, the parameter $\alpha_{\rm H}^{(0)}$ should be 
less than the order of $10^{-10}$ \cite{Creminelli:2018xsv}. 
As we mentioned in Introduction, it is still a matter of debate whether 
the EFT of dark energy is valid around the frequency 
$f \sim 100$~Hz \cite{deRham:2018red}.
In this paper, we will not impose such a bound and independently test how the 
cosmological observations place the upper limit of $ x_4^{(0)}$.
 
In Fig.~\ref{fig:stability}, we show the physically viable parameter space (blue colored region) 
for the initial conditions $x_1^{(s)}$, $x_2^{(s)}$, $x_3^{(s)}$, $x_4^{(s)}$ 
(at redshift $z_s=1.5 \times 10^{5}$) and today's values 
$x_1^{(0)}, x_2^{(0)}, x_3^{(0)}$ (at redshift $z=0$). 
We find that $x_1^{(0)}$ is negative, while $x_2^{(0)}$ and $x_3^{(0)}$ are positive. 
We note that the ghost condensate model \cite{Arkani} 
has a de Sitter solution satisfying $x_1<0$ and $x_2>0$. 
The Galileon term $x_3$ modifies the cosmological dynamics of ghost condensate,  
but there is also a de Sitter attractor characterized by $x_1<0$, 
$x_2>0$, and $x_3>0$ \cite{Kase:2018iwp}. 
As we see in Fig.~\ref{fig:stability}, the parameter $x_3^{(0)}$ is 
not well constrained from the theoretically viable conditions alone.

The parameter space of the variable $x_4^{(0)}$ is not shown in Fig.~\ref{fig:stability}, 
but it is in the range $|x_4^{(0)}| \ll 1$ to satisfy all the theoretically consistent 
conditions. As $x_4^{(0)}$ approaches the order 1, the scalar perturbation is 
typically prone to the Laplacian instability associated with 
the negative value of $c_s^2$ \cite{Kase:2018iwp}.

The above results will be used to set theoretical priors for the MCMC analysis.

\section{Cosmological perturbations }\label{Sec:perturbations}

In this section, we discuss the evolution of scalar cosmological perturbations 
in the BH model for the perturbed line element given by Eq.~(\ref{permet}).
We introduce the two gauge-invariant gravitational potentials:
\be
\Psi \equiv \delta N+\dot{\psi}\,,\qquad 
\Phi \equiv -\zeta-H\psi\,.
\label{grapo}
\ee
For the matter sector, we consider scalar perturbations of the matter-energy momentum 
tensor $T^{\mu}_{\nu}$ arising from the action ${\cal S}_M$, as 
$\delta T^0_0=-\delta \rho$, $\delta T^0_i=\partial_i \delta q$, and 
$\delta T^i_j=\delta P \delta^i_j$. 
The density perturbation $\delta \rho$, the momentum perturbation $\delta q$, 
and the pressure perturbation $\delta P$ are expressed in terms of the 
sum of each matter component, as 
$\delta \rho=\sum_{i} \delta \rho_i$, 
$\delta q=\sum_{i} \delta q_i$, and 
$\delta P=\sum_{i} \delta P_i$, where $i=m,r$.
We introduce the gauge-invariant density contrast:
\be
\Delta_i \equiv \frac{\delta \rho_i}{\rho_i}
-3H \frac{\delta q_i}{\rho_i}\,,
\ee
where $\rho_i$ is the background density of each component.
In the BH model, the full linear perturbation equations of motion 
were derived in Ref.~\cite{Kase:2018iwp}.

In Fourier space with the comoving wavenumber $k$, 
we relate the gravitational potentials in Eq.~(\ref{grapo}) 
with the total matter density contrast $\Delta=\sum_{i} \Delta_i$, 
as \cite{Amendola:2007rr,Bertschinger:2008zb,Pogosian:2010tj}
\ba 
-k^2\Psi &=&4\pi G_{\rm N} a^2\mu(a,k)\rho\Delta\,, \label{gra1}\\
-k^2(\Psi+\Phi)&=&8\pi G_{\rm N} a^2\Sigma(a,k)\rho\Delta\,,\label{gra2}
\ea
where $G_{\rm N}$ is the Newton gravitational constant given by 
Eq.~(\ref{GN}), and $\rho=\sum_{i} \rho_i$ is the total background matter density.  
The dimensionless quantities $\mu$ and $\Sigma$ correspond to 
the effective gravitational couplings felt by matter and light, respectively.
For nonrelativistic matter, the density contrast $\Delta_m$ obeys \cite{Kase:2018iwp}
\be
\ddot{\Delta}_m+2H \dot{\Delta}_m+\frac{k^2}{a^2} \Psi
=-3 \left( \ddot{\cal B}+2H \dot{\cal B} 
\right)\,,
\ee
where ${\cal B} \equiv \zeta+H \delta q_m/\rho_m$. 
This means that the matter density contrast grows due to 
the gravitational instability through the modified Poisson 
Eq.~(\ref{gra1}).
In GR, both $\mu$ and $\Sigma$ are equivalent to 1, but 
in the BH model, they are different from 1. 
Hence the growth of structures and gravitational potentials 
is subject to modifications.

For the perturbations deep inside the sound horizon ($c_s^2k^2/a^2 \gg H^2$), the common 
procedure is to resort to a quasi-static approximation for the estimations 
of  $\mu$ and $\Sigma$ \cite{Pola00,Tsujikawa:2007gd,DKT}. 
This amounts to picking up the terms containing $k^2/a^2$ and $\Delta_m$ in the perturbation 
equations of motion. In Horndeski theories, it is possible to obtain the closed form 
expressions of $\Psi, \Phi, \zeta$ \cite{DKT,Kase:2018aps}. 
In GLPV theories, the additional time derivatives $\alpha_{\rm H} \dot{\psi}$ and 
$\alpha_{\rm H} \dot{\zeta}$ appear even 
under the quasi-static approximation \cite{DeFelice:2015isa,Tsujikawa:2015mga}, 
so the perturbation equations are not closed. 
If $|\alpha_{\rm H}|$ is very much smaller than 1 
and $x_4$ is subdominant to $x_{1,2,3}$, 
we may ignore the contributions of the term $x_4$ to the perturbation equations. 
In this case, we can estimate $\mu$ and $\Sigma$ 
in the BH model, as \cite{Kase:2018iwp}
\be
\mu \simeq \Sigma \simeq 
1+\frac{2Q_t x_3^2}{Q_s c_s^2 (2-x_3)^2}\,.
\label{musi}
\ee
Since $\mu$ and $\Sigma$ are identical to each other, it follows that 
$\Psi \simeq \Phi$. Under the theoretically consistent conditions 
(\ref{Qt}), (\ref{Qs}), and (\ref{cs}), we also have $\mu \simeq \Sigma>1$ 
and hence the gravitational interaction is stronger than that in GR.
Let us note that in the following we will not rely on this approximation 
and we will solve the complete linear perturbation equations. 

\begin{table}
\begin{center}
\begin{tabular}{|c|c|c|c|c|}
\hline
Parameters & BH1 & BH2 & BH3 & GGC \\
\hline
\hline
$x_{1}^{(s)} \, \, (\cdot 10^{-16})$&$-1$&$-0.1$ & $-0.01$ & $-1$   \\
$x_{2}^{(s)}\, \, (\cdot 10^{-16})$ &$5$& $0.05$  &  $0.0001$ & $5$  \\
$x_{3}^{(s)} \, \, (\cdot 10^{-9})$ &$1$& $1$ & $ 0.1$ &  $10$  \\
$x_{4}^{(s)} \, \, (\cdot 10^{-6})$&$100$ & $1$ & $1$ & $0$ \\
\hline
$x_{1}^{(0)}$ &$-1.37$& -1.03 &$-0.73$ & -1.23 \\
$x_{2}^{(0)}$ &$2.03$& 1.02 & $0.12$ &   1.63 \\
$x_{3}^{(0)}$ &$0.03$& 0.69 &$ 1.30 $ &  0.29   \\
$x_{4}^{(0)}$ &$1\cdot 10^{-5}$& $5\cdot 10^{-6}$ & $2 \cdot 10^{-4} $ & $ 0 $ \\
\hline
\end{tabular}
\caption{List of starting values of the density parameters $x_i$
at the redshift $z_s=1.5 \times 10^{5}$ and corresponding 
today's values for three  beyond-Horndeski (BH) models and the f Galileon Ghost Condensate (GGC) model with $x_4=0$.   The BH1, BH2 and BH3 models  differ in the starting values $x_i^{(s)}$.
All of them satisfy theoretically consistent conditions discussed 
in Sec.~\ref{Sec:perturbations}. 
We study these models for the purpose of visualizing and 
quantifying the modifications from $\Lambda$CDM. 
 The cosmological parameters (e.g., $H_0, \Omega_m, \Omega_r$) used for 
these models are the Planck 2015 best-fit values for $\Lambda$CDM \cite{Ade:2015xua}. }
\label{table:parameters}
\end{center}
\end{table}

\begin{figure}
\includegraphics[width=.49\textwidth]{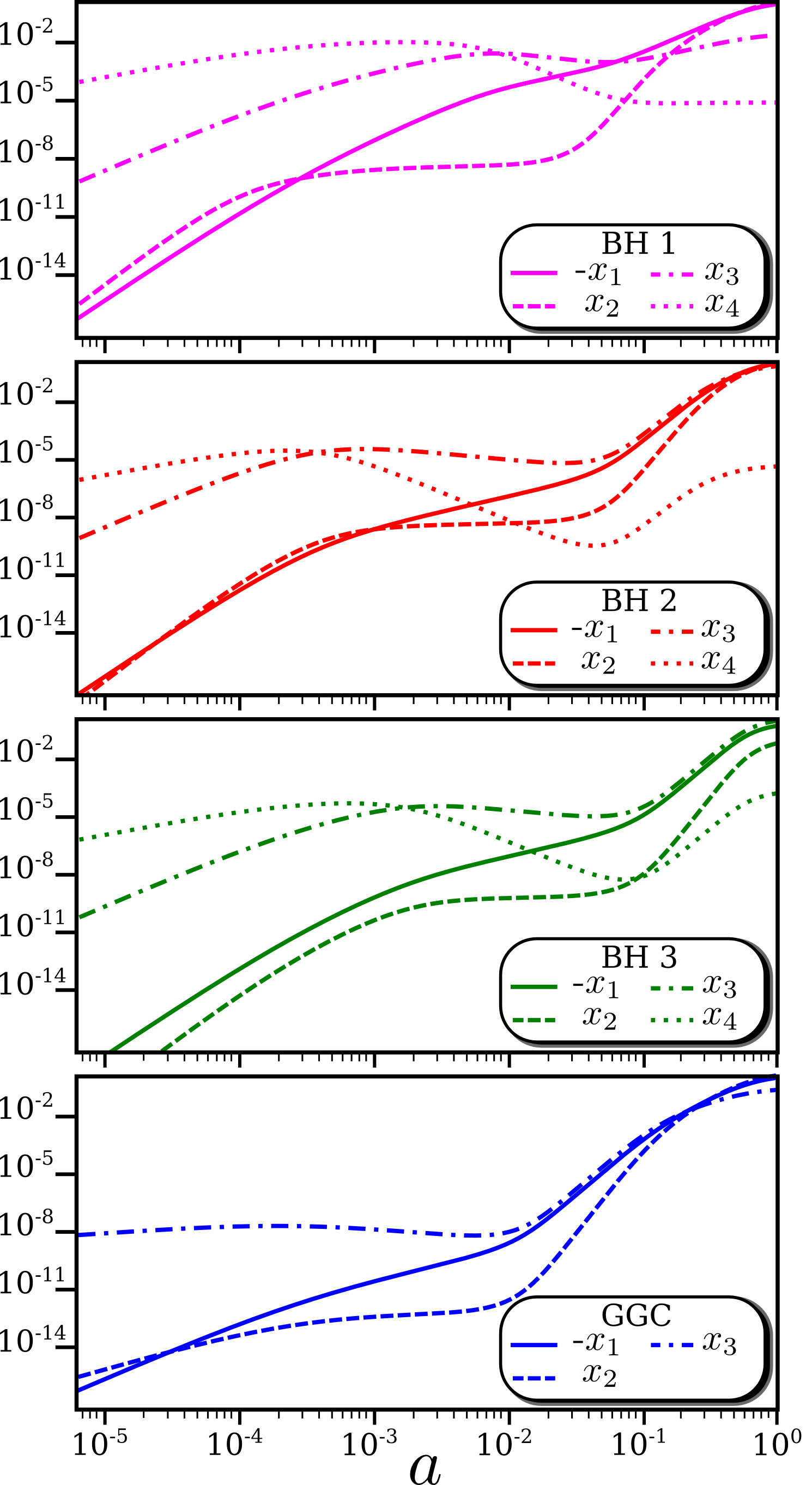}\\
\caption{Evolution of the dimensionless variables defined 
in Eq.~(\ref{dimensionless_functions}) versus the scale 
factor $a$ (with today's value $1$) for four test models 
listed in Table~\ref{table:parameters}. 
 In this Table, the staring values of parameters $x_i$ 
at the initial redshift $z_s=1.5 \times 10^{5}$ are shown 
for each test model. 
We discuss physical implications for the evolutions of $x_i$  
in Sec.~\ref{Sec:perturbations}.
\label{fig:xi}}
\end{figure}

To understand the evolution of perturbations, we consider four different 
cases (BH1, BH2, BH3, GGC) listed in Table \ref{table:parameters}. 
The difference between these models is characterized by the different 
choices of initial conditions $x_i^{(s)}$  at the redshift $z_s=1.5 \times 10^{5}$.
Among them, BH1 has the largest initial value $x_4^{(s)}$, 
while $x_4$ is always 0 in GGC (which belongs to Horndeski theories).
In Fig.~\ref{fig:xi}, we plot the evolution of $x_i$ from the past to today
for these four different cases.  
In BH1, the variable $x_4$ dominates over other variables $x_{1,2,3}$ 
for $a \lesssim 10^{-2}$, but it becomes subdominant at low redshifts 
with today's value of order $10^{-5}$. 
Comparing BH1 with BH3, we observe that the initial largeness of 
$x_4$ does not necessarily imply the large present-day value $x_4^{(0)}$. 
At low redshifts, $x_4$ is typically less than the order $10^{-3}$ to avoid 
$c_s^2<0$ with the amplitude smaller than $x_{1,2,3}$, 
in which case the analytic estimation (\ref{musi}) can be trustable.
Indeed, for all the models given in Table \ref{table:parameters}, 
we numerically checked that the quasi-static approximation 
holds with sub-percent precision for the wavenumbers $k>0.01$ Mpc$^{-1}$ 
(as confirmed in Horndeski theories in Ref.~\cite{Peirone:2017ywi, Frusciante:2018jzw}).

\begin{figure}
\includegraphics[width=.49\textwidth]{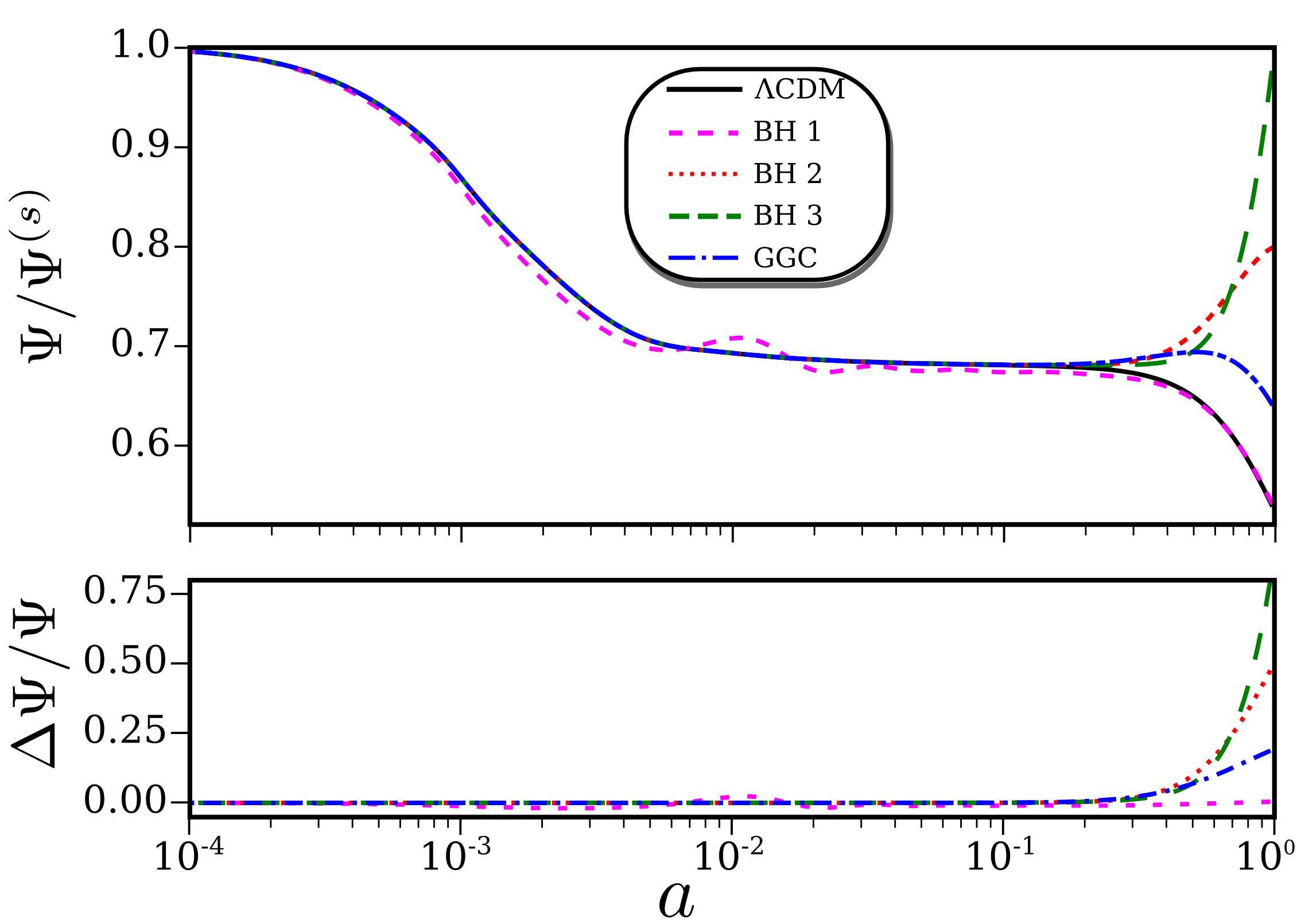}
\caption{(Top) Evolution of the gravitational potential $\Psi$ normalized 
by its initial value $\Psi^{(s)}$ for the wavenumber $k = 0.01$~Mpc$^{-1}$. 
We show the evolution of $\Psi/\Psi^{(s)}$ for four models listed in 
Table~\ref{table:parameters} and also for $\Lambda$CDM (black line).  
(Bottom) Percentage relative difference of $\Psi$ relative to that in 
$\Lambda$CDM. The cosmological parameters used for this plot are 
the Planck 2015 best-fit values for 
$\Lambda$CDM \cite{Ade:2015xua} (which is also the case for 
plots in Figs.~\ref{fig:lensing_spectra} and \ref{fig:ISW}). 
 The physical interpretation of this figure is discussed 
in Sec.~\ref{Sec:perturbations}.
\label{fig:potentials} 
}
\end{figure}

In the top panel of Fig.~\ref{fig:potentials}, we plot the evolution of $\Psi$ normalized 
by its initial value $\Psi^{(s)}$ for the four models in Table \ref{table:parameters} and 
for the $\Lambda$CDM. In the bottom panel, we depict the percentage difference of $\Psi$ for the chosen models with respect to $\Lambda$CDM. At the late epoch, the deviations from 
$\Lambda$CDM show up with the enhanced gravitational potential 
(around $a \sim 0.2$  for the BH2, BH3, GGC models). 
The largest deviation arises for BH3, in which case the difference is more than 75\,\% today. 
As estimated from Eq.~(\ref{musi}), the modified evolution of $\Psi$ is mostly attributed to 
the cubic Galileon term $x_3$. 
For larger today's values of $x_3^{(0)}$, the difference of $\Psi$ from $\Lambda$CDM 
tends to be more significant with the larger deviation of $\mu$ from 1.
In Fig.~\ref{fig:potentials}, we observe that the deviation from $\Lambda$CDM 
increases with the order of BH1, GGC, BH2, BH3, by reflecting their increasing
values of $x_3^{(0)}$ given in Table \ref{table:parameters}.

\begin{figure}
\includegraphics[width=.49\textwidth]{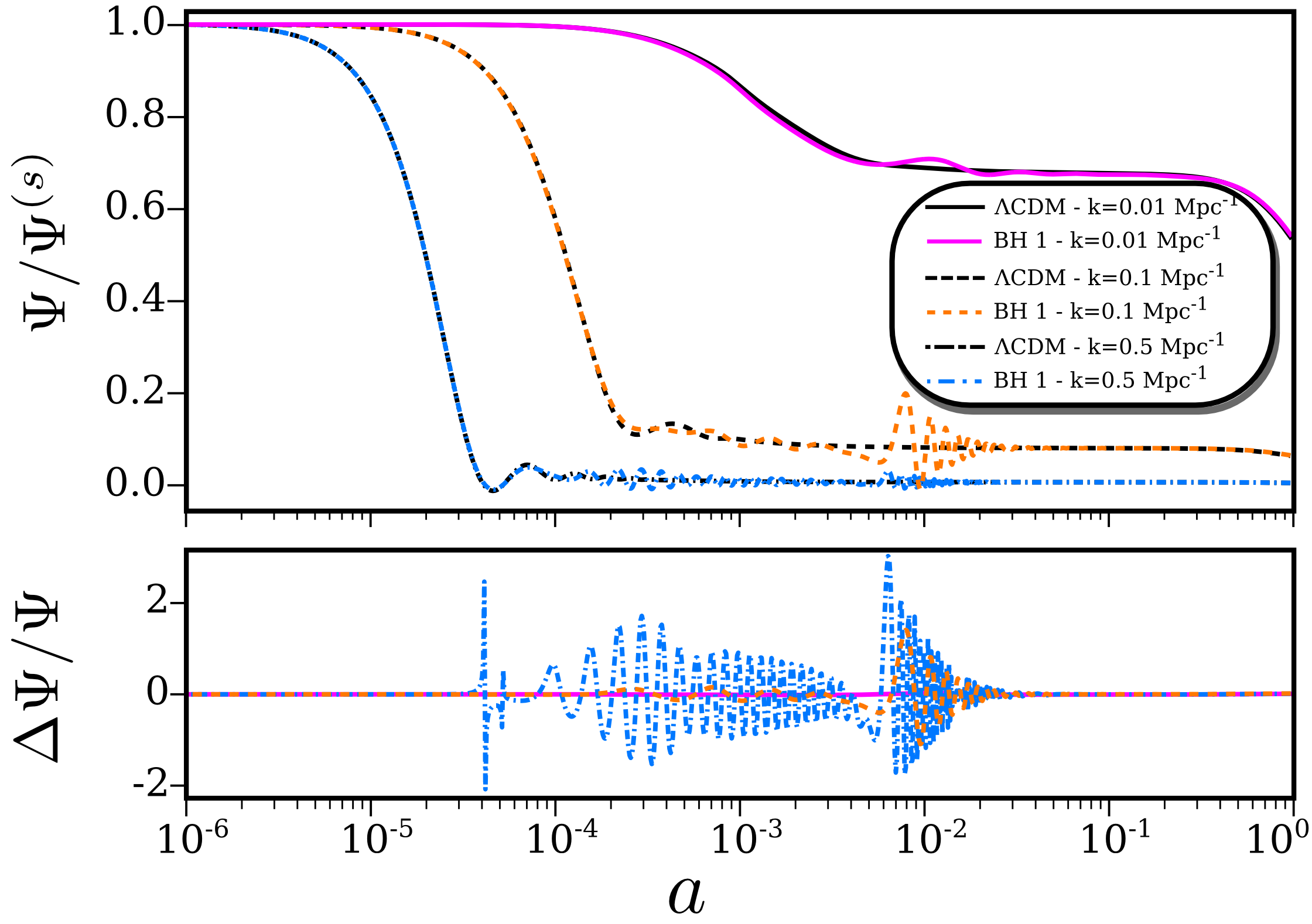}
\caption{(Top) Evolution of the gravitational potential $\Psi$ normalized 
by its initial value $\Psi^{(s)}$ for BH1 and $\Lambda$CDM
with three different wavenumbers: $k = 0.01, 0.1, 0.5$~\,Mpc$^{-1}$.  
 In Table~\ref{table:parameters}, we list the starting values of 
parameters $x_i$ at the initial redshift $z_s=1.5 \times 10^{5}$ for the BH1 model.
(Bottom) Percentage relative difference of $\Psi$ relative to that 
in $\Lambda$CDM for the same values of $k$ in the top panel. 
\label{fig:BH1k} 
}
\end{figure}

In BH1, there is the suppression of $|\Psi|$ in comparison to $\Lambda$CDM 
at high redshifts ($a \lesssim 10^{-2}$). 
This property arises from the dominance of $x_4$ over $x_{1,2,3}$ at early times, 
in which case the relative density abundances 
between dark energy and matter fluids are modified.
Besides this effect, the non-negligible early-time contribution of $x_4$ to 
scalar perturbations gives rise to a scale-dependent evolution of  gravitational 
potentials, which manifests itself in the $k$-dependent variation of $\mu(a,k)$ and $\Sigma(a,k)$.
In Fig.~\ref{fig:BH1k}, we plot the evolution of $\Psi$ in BH1 
for three different values of $k$.
For perturbations on smaller scales, the deviation from $\Lambda$CDM 
tends to be more significant.
In models BH2, BH3, and GGC, the early-time evolution of $\Psi$ is similar to 
that in $\Lambda$CDM, but they exhibit large deviations from $\Lambda$CDM
at late times.

\begin{figure}
\includegraphics[width=.48\textwidth]{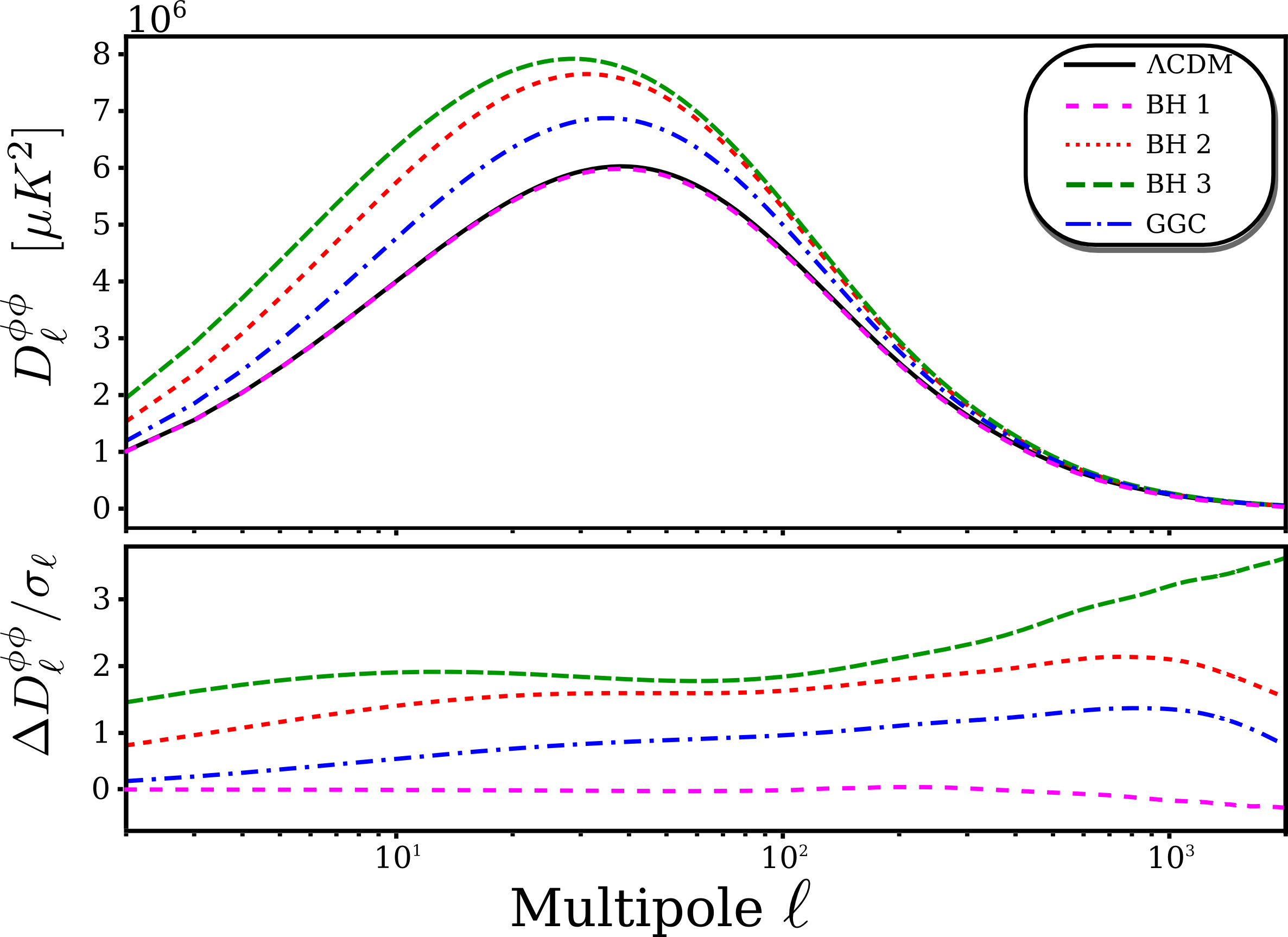}
\caption{(Top) Lensing angular power spectra 
$D_\ell^{\phi \phi} = \ell(\ell+1)C_\ell^{\phi \phi}/(2 \pi)$ 
for $\Lambda$CDM and the models listed 
in Table~\ref{table:parameters},  where $C_\ell$ is 
defined by Eq.~(\ref{phiphispectrum}).
(Bottom) Relative difference of the lensing angular 
power spectra, computed with respect to $\Lambda$CDM, 
in units of the cosmic variance 
$\sigma_\ell = \sqrt{2/(2 \ell+1)} C_\ell^{\Lambda {\rm CDM}}$.
\label{fig:lensing_spectra} 
}
\end{figure}

\begin{figure}
\includegraphics[width=.48\textwidth]{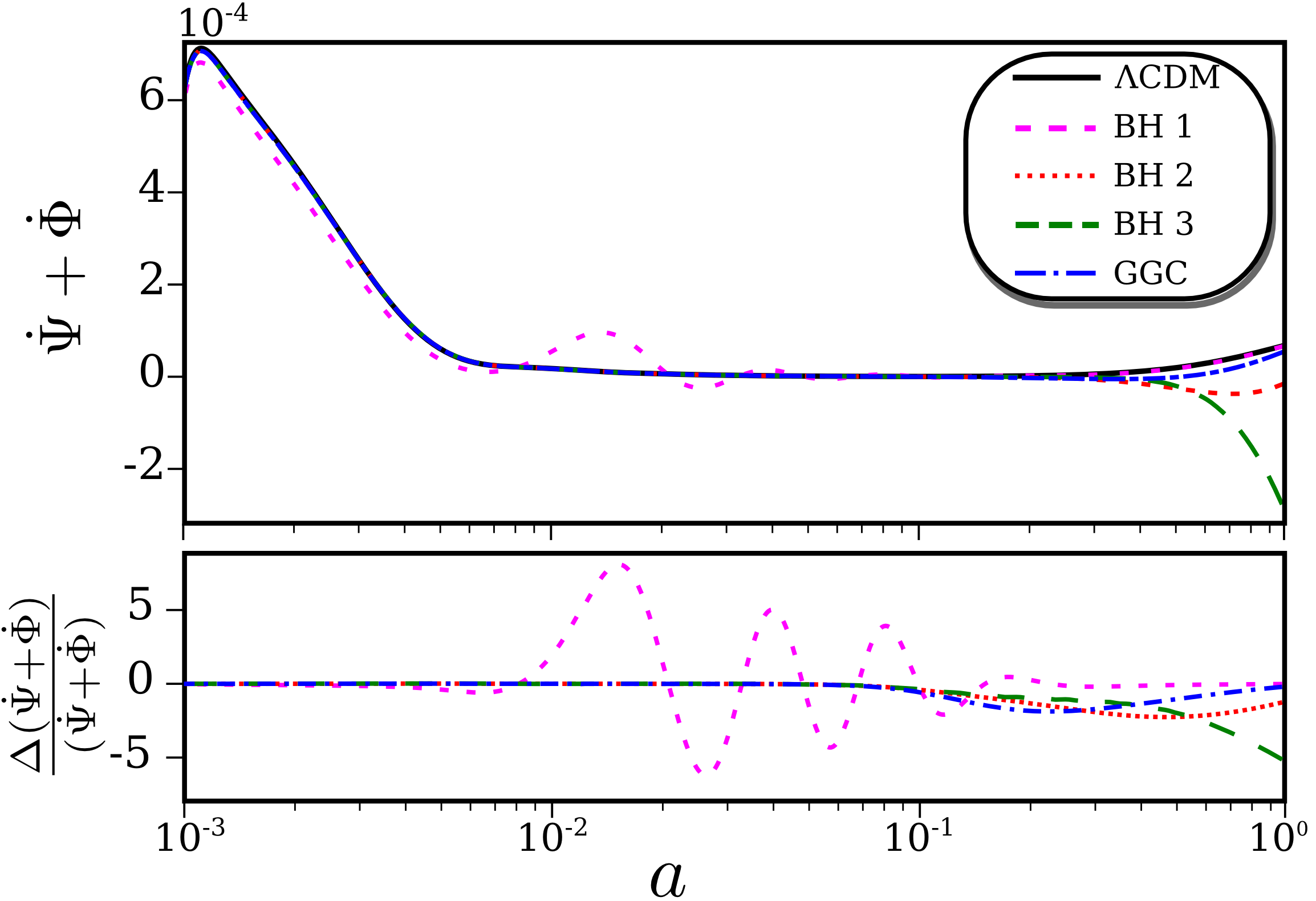}
\caption{(Top) Evolution of the time derivative 
$\dot \Psi +\dot \Phi$ 
for $\Lambda$CDM and the models listed 
in Table~\ref{table:parameters}, computed at $k = 0.01$~Mpc$^{-1}$.
(Bottom) Relative difference of $\dot \Psi +\dot \Phi$, 
computed with respect to $\Lambda$CDM. 
 See the discussion after Eq.~(\ref{T_source}) for the 
physical interpretation of this figure.
\label{fig:ISW} 
}
\end{figure}

At low redshifts, the lensing gravitational potential 
$\phi_{\rm len}=(\Psi+\Phi)/2$ evolves in a similar way to $\Psi$, 
by reflecting the property $\mu \simeq \Sigma$ for $x_4^{(0)} \ll 1$.
The lensing  angular power spectrum can be computed by using 
the line of sight integration method, with 
the convention \cite{Lewis:2006fu}
\begin{align}
\label{phiphispectrum}
C_\ell^{\phi\phi}=4\pi\int \frac{{\rm d}k}{k}\mathcal{P}(k)
\left[\int_0^{\chi_{\ast}}{\rm d}\chi\,
S_{\phi}(k;\tau_0-\chi)j_\ell(k\chi)\right]^2\,,
\end{align}
where $\mathcal{P}(k)=\Delta^2_\mathcal{R}(k)$ is the primordial power spectrum of curvature perturbations, and $j_{\ell}$ 
is the spherical Bessel function.
The source $S_{\phi}$ is expressed in terms of the transfer function 
\be
S_{\phi}(k;\tau_0-\chi)=2T_{\phi}(k;\tau_0-\chi)\left(\frac{\chi_{\ast}-\chi}{\chi_{\ast}\chi}\right)\,, 
\ee
with $T_{\phi}(k,\tau) = k\phi_{\rm len}$, $\chi$ is the comoving 
distance with $\chi_{\ast}$ corresponding to that to the last scattering surface, $\tau_0$ is today's conformal time $\tau=\int a^{-1} \rd t$ 
satisfying the relation $\chi=\tau_0-\tau$.
In Fig.~\ref{fig:lensing_spectra}, we show the lensing power spectra 
$D_\ell^{\phi \phi}=\ell(\ell+1)C_\ell^{\phi \phi}/(2 \pi)$ and relative differences in units of the cosmic variance 
for four models listed in Table \ref{table:parameters}. 
Since $\Sigma>1$ at low redshifts in BH and GGC models, this
works to enhance $D_\ell^{\phi \phi}$ compared to $\Lambda$CDM. 
We note that the amplitude of matter density contrast $\delta_m$ 
in these models also gets larger than that in $\Lambda$CDM 
by reflecting the fact that $\mu>1$. 
In Fig.~\ref{fig:lensing_spectra}, we observe that, apart from BH1 in which $\Sigma$ is close to 1, the lensing power spectra in other three cases are subject to the enhancement with respect to $\Lambda$CDM. 
Since today's values of $\mu$ and $\Sigma$ increase for larger $x_3^{(0)}$, 
the deviation from $\Lambda$CDM tends to be more significant with 
the order of GGC, BH2, and BH3.

\begin{figure*}
\includegraphics[height=4.3in,width=6.8in]{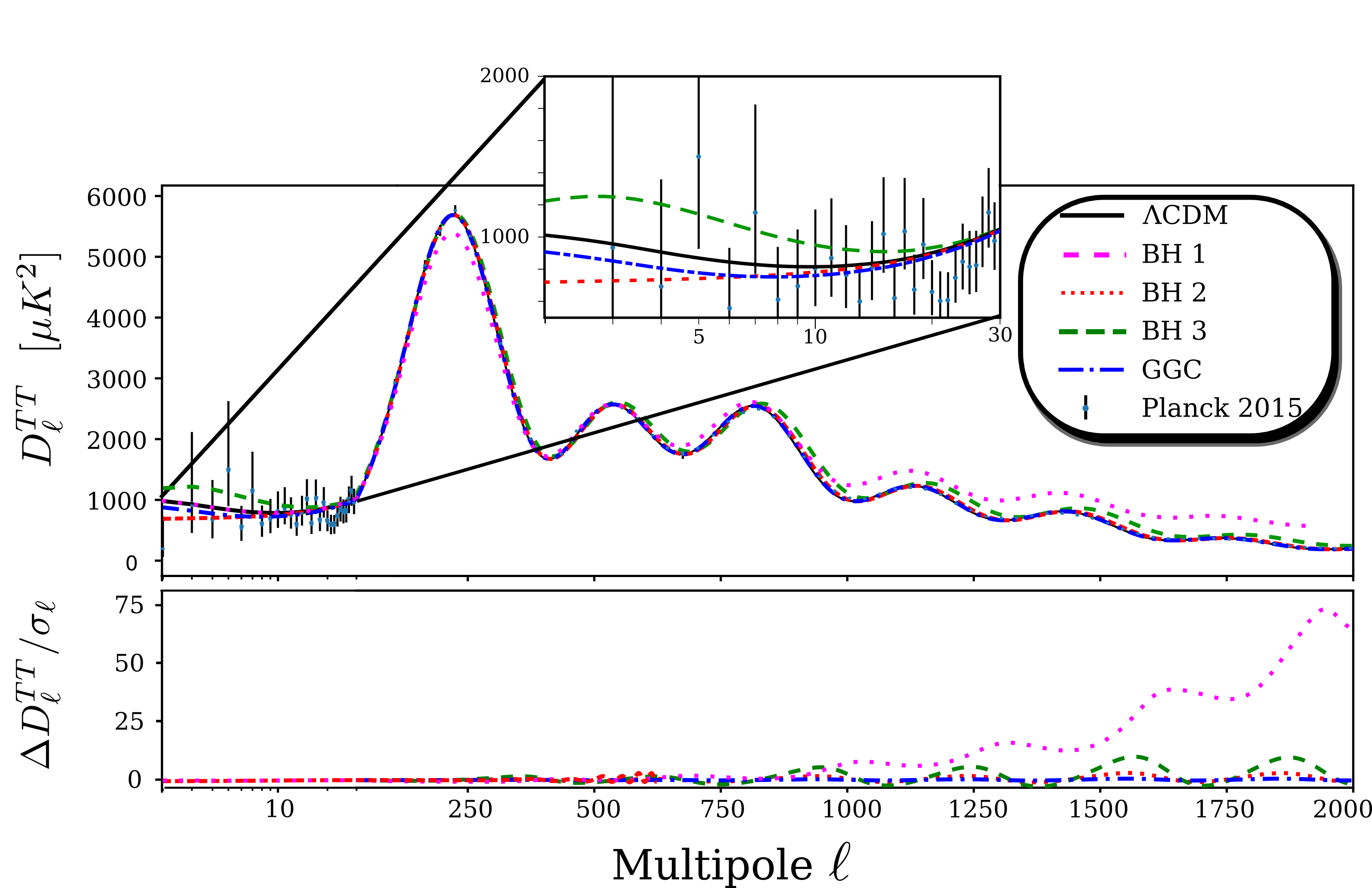}
\caption{(Top) CMB TT power spectra 
$D_\ell^{\rm TT} = \ell(\ell+1)C_\ell^{\rm TT}/(2 \pi)$ 
for the  test models presented in Table~\ref{table:parameters}, 
compared with data points from the Planck 2015 release. 
(Bottom) Relative difference of TT power spectra, computed with 
respect to $\Lambda$CDM in units of the cosmic variance 
$\sigma_\ell = \sqrt{2/(2 \ell+1)} C_\ell^{\Lambda {\rm CDM}}$. 
\label{fig:TT_spectra} 
}
\end{figure*}

\begin{figure}
\includegraphics[width=.49\textwidth]{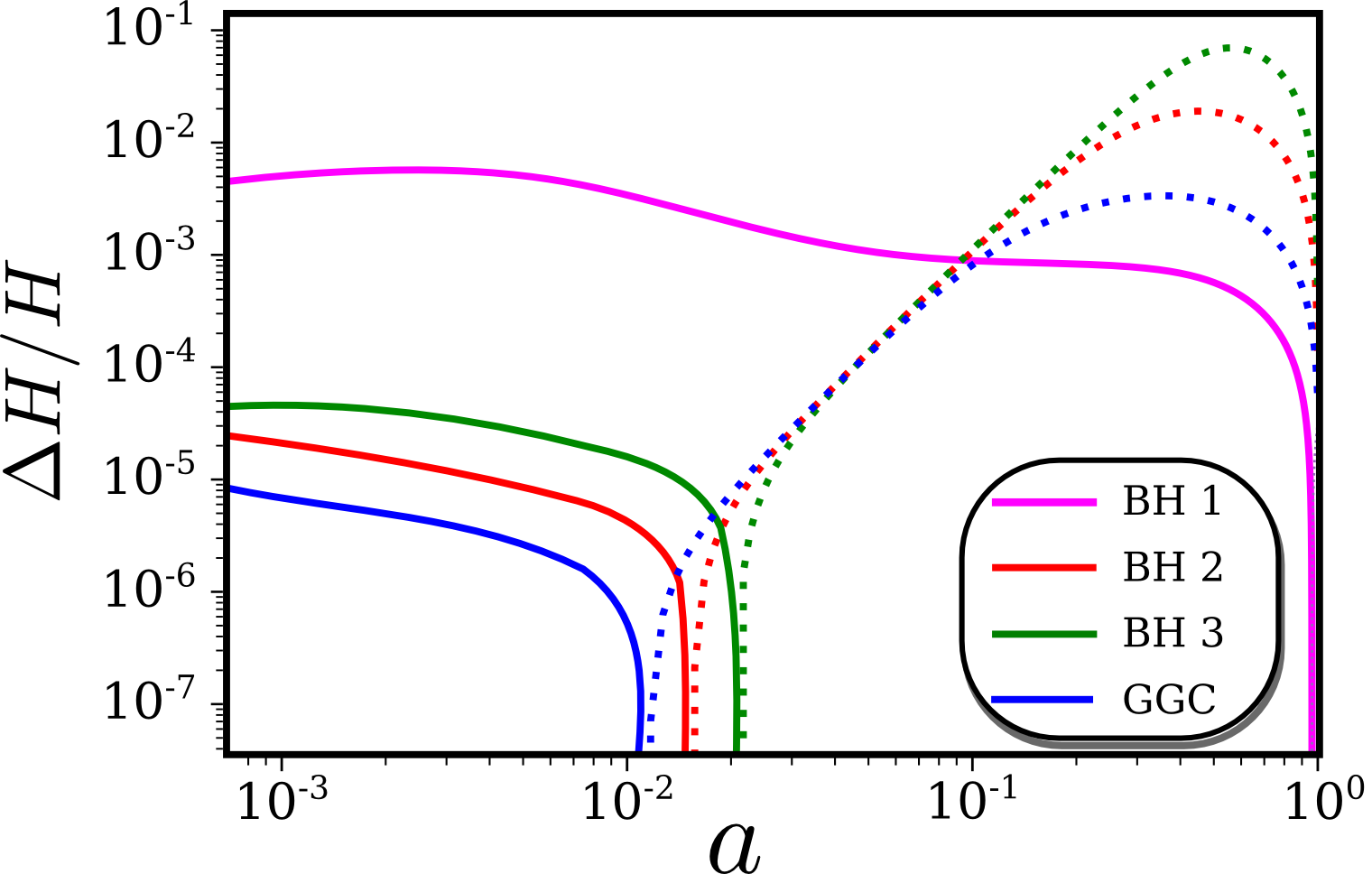}
\caption{Evolution of the relative Hubble rate for  the models listed in 
Table \ref{table:parameters} compared to $\Lambda$CDM. 
The solid lines correspond to a positive difference, whereas 
the opposite holds for the dashed lines. 
 For BH1 the largest difference from 
$\Lambda$CDM occurs in the early cosmological epoch, 
in which case the CMB acoustic peaks shift 
toward lower multipoles.
\label{fig:hubblerate} 
}
\end{figure}

Let us proceed to the discussion of the impact of BH and GGC models on 
the CMB temperature anisotropies.   The CMB temperature-temperature (TT) angular spectrum can be expressed as~\cite{Seljak:1996is}
\be
\label{TTspectrum}
C_\ell^{\rm TT}=
(4\pi)^2\int \frac{{\rm d}k}{k}~\mathcal P(k)\Big|\Delta_\ell^{\rm T}(k)\Big|^2\,,
\ee
where 
\be
\Delta_\ell^{\rm T}(k)=
\int_0^{\tau_0}{\rm d}\tau\,
e^{ik\tilde{\mu}(\tau-\tau_0)}S_{\rm T}(k,\tau)
j_\ell[k(\tau_0-\tau)]\,,
\ee
with $\tilde{\mu}$ being the angular separation, and 
$S_{\rm T}(k,\tau)$ is the radiation transfer function.
The contribution to $S_{\rm T}(k,\tau)$ arising from the 
integrated-Sachs-Wolfe (ISW) effect is of the form
\begin{align}
\label{T_source}
S_{\rm T}(k,\tau) \sim \left(\frac{\rd \Psi}{\rd \tau}+\frac{\rd \Phi}{\rd \tau} \right)
e^{-\kappa}\,, 
\end{align}
where $\kappa$ is the optical depth. 
Besides the early ISW effect which occurs during the transition from 
the radiation to matter eras by the time variation of 
$\Psi+\Phi$, the presence of dark energy 
induces the late-time ISW effect. 
In the $\Lambda$CDM model, the gravitational potential $-(\Psi+\Phi)$, 
which is positive, decreases by today with at least more than 30 \% relative to its initial value 
(see Fig.~\ref{fig:potentials}). 
As we observe in Fig.~\ref{fig:ISW} we have $\dot{\Psi}+\dot{\Phi}>0$ in this case, so 
the ISW effect gives rise to the positive contribution to  Eq.~(\ref{TTspectrum}).
In Fig.~\ref{fig:TT_spectra}, we plot the CMB TT power spectra 
$D_{\ell}^{\rm TT}=\ell (\ell+1)C_{\ell}^{\rm TT}/(2\pi)$ for
the models listed in Table \ref{table:parameters} and 
$\Lambda$CDM. 
In BH1 the parameter $\Sigma$ is close to 1 at low redshifts due to the smallness of $x_3^{(0)}$, so the late-time ISW effect works in the similar way to the GR case. 
Hence the TT power spectrum in BH1 for the multipoles $\ell \lesssim 30$ 
is similar to that in $\Lambda$CDM. 

In the GGC model of Fig.~\ref{fig:TT_spectra}, we observe that the large-scale ISW tail is suppressed relative to that 
in $\Lambda$CDM. This reflects the fact that the larger deviation of 
$\Sigma$ from 1 leads to the time derivative $\dot{\Psi}+\dot{\Phi}$ closer to 0, 
see Fig.~\ref{fig:ISW}. Hence the late-time ISW effect is not significant, 
which results in the suppression of $D_{\ell}^{\rm TT}$ with respect to $\Lambda$CDM. 
In Ref.~\cite{PBFT} this fact was first recognized in the GGC model, which exhibits a better 
fit to the Planck CMB data. 
As the deviation of $\Sigma$ from 1 increases further, the sign of $\dot{\Psi}+\dot{\Phi}$ 
changes to be negative (see Fig.~\ref{fig:ISW}). The BH2 model can be regarded as such a marginal case in which 
the large-scale ISW tail is nearly flat. 
In BH3, the increase of $\Sigma$ at low redshifts is so significant that the largely 
negative ISW contribution to Eq. (\ref{TTspectrum}) leads to the enhanced 
low-$\ell$ TT power spectrum relative to $\Lambda$CDM. 

The modified evolution of the Hubble expansion rate from $\Lambda$CDM 
generally leads to the shift of CMB acoustic peaks at high-$\ell$. 
In Fig.~\ref{fig:hubblerate}, we observe that the largest deviation 
of $H(a)$  at high redshifts occurs for BH1 by the dominance of $x_4$ over $x_{1,2,3}$.
This leads to the shift of acoustic peaks toward lower multipoles (see Fig.~\ref{fig:TT_spectra}). 
We also find that BH3 is subject to non-negligible shifts of high-$\ell$ peaks due to the large modification of $H(a)$ at low redshifts, 
in which case the peaks shift toward higher multipoles.
Moreover, there is the large enhancement of ISW tails for BH3, 
so it should be tightly constrained from the CMB data. 
We note that the shift of CMB acoustic peaks is further constrained 
by the datasets of BAO and SN~Ia.
For BH2 and GGC the changes of peak positions 
are small in comparion to BH1 and BH3, but still they are in the range 
testable by the CMB data.
Moreover, the large-scale ISW tail is subject to the suppression relative to
$\Lambda$CDM in BH2 and GGC.

In BH1, we also notice a change in the amplitude of acoustic peaks 
occurring dominantly at high $\ell$. 
This is known to be present in models with early-time 
modifications of gravity \cite{Xiang-Raveri,Benevento2018}. 
The modification of gravitational potentials affects the evolution of radiation perturbations (monopole and dipole) through the radiation driving effect \cite{Hu:1994uz,Xiang-Raveri}, thus resulting in the changes in amplitude and phase of acoustic peaks 
at high $\ell$. 

The modified time variations of $\Psi$ and $\Phi$ around the 
recombination epoch also give a contribution to the early ISW effect. 
This is important on scales around the first acoustic peak, corresponding to 
the wavenumber $k \simeq 0.016$~Mpc$^{-1}$ for our choice of model parameters. 
To have a more qualitative feeling of this contribution, we have estimated the impact of the early ISW effect on $D_{\ell}^{\rm TT}$ by using the approximate ISW integral presented 
in Ref.~\cite{Hu:1994uz}:
\be
\int_{\tau_*}^{\tau_0} \rd \tau  \left( 
\frac{\rd \Psi}{\rd \tau}+\frac{\rd \Phi}{\rd \tau} \right) j_\ell
 \left[ k (\tau_0-\tau) \right]
\simeq \left[ \Psi+\Phi \right] |_{\tau_*}^{\tau_0} j_\ell (k\tau_0)\,,
\ee
where $\tau_*$ is the conformal time at the last scattering.
Then, we find a negative difference of about 4.9\,\% between BH1 and $\Lambda$CDM. 
This is in perfect agreement with the change in amplitude of the first acoustic peak 
shown in Fig.~\ref{fig:TT_spectra}. 
Thus, the BH models in which $x_4$ is the dominant contribution to the dark energy 
dynamics at early times can be severely constrained from the CMB data. 

We stress that, in the late Universe, $x_4$ is typically suppressed compared to
$x_{1,2,3}$ for the viable cosmological background, so the main impact on 
the evolution of perturbations comes from the cubic Galileon term $x_3$.
The analytic estimation (\ref{musi}) is sufficiently trustable for studying the evolution of gravitational potentials and matter perturbations at low redshifts.
However, we solve the full perturbation equations of motion for the 
MCMC analysis without resorting to the quasi-static approximation.

\section{Observational constraints}\label{Sec:constraints}

We place observational bounds on the BH model by performing the MCMC 
simulation with different combinations of datasets at high and low redshifts.

\subsection{Datasets}
\label{Sec:datasets}

For the MCMC likelihood analysis, based on the EFTCosmoMC code, we use the Planck 2015~\citep{Aghanim:2015xee,Ade:2015xua} 
data of CMB temperature and polarization on large angular scales, for multipoles 
$\ell<29$ (low-$\ell$ TEB likelihood) 
and the CMB temperature on smaller angular scales (PLIK TT Likelihood).
We also consider the BAO measurements from the 6dF Galaxy Survey~\citep{BAO1} 
and  from the SDSS DR7 Main Galaxy Sample~\citep{BAO2}. 
Moreover, we include the combined BAO and RSD datasets from the SDSS DR12 
consensus release \citep{Alam:2016hwk} and  the JLA SN Ia 
sample \citep{Betoule:2014frx}. We will refer to the full combined datasets as ``PBRS''. 

Finally, we impose the flat priors on the model parameters: 
$x_1^{(s)} \in [-10, 10] \times 10^{-16}$, 
$x_3^{(s)} \in [-10, 10]\times 10^{-9}$, and 
$x_4^{(s)} \in [0, 10]\times 10^{-6}$. 
Even by increasing the prior volume by one order of magnitude, 
we confirmed that the likelihood results are not subject to the priors choice.

\subsection{Constrained parameter space}
\label{Sec:results}

In this section, we show observational constraints on model parameters in the BH model. 
We use the datasets presented in Sec.~\ref{Sec:datasets} with two combinations: 
(i) Planck and (ii) PBRS. For reference, we also present the results 
of the $\Lambda$CDM model.

\begin{figure}[t!]
\includegraphics[width=.37\textwidth]{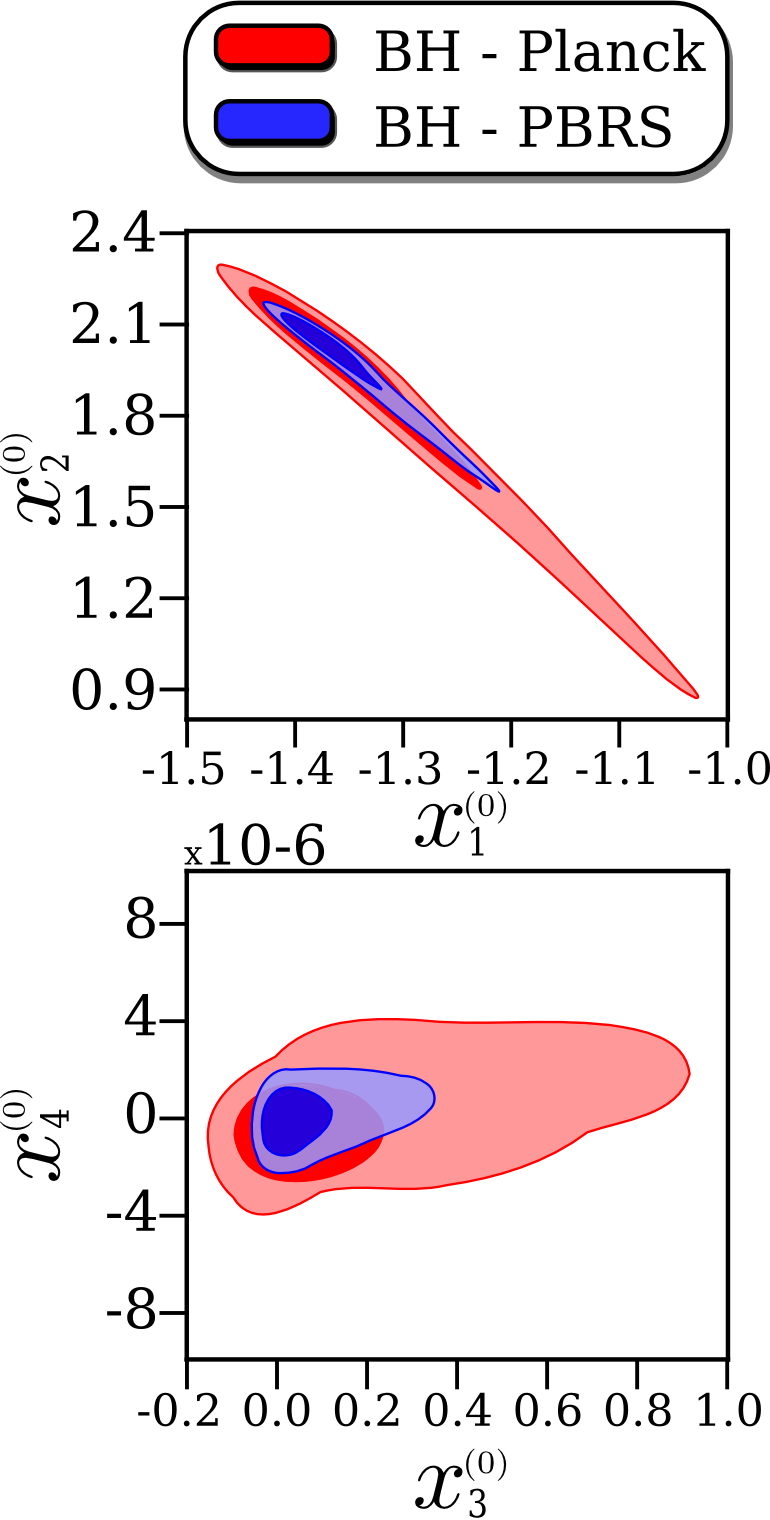}
\caption{Two-dimensional observational bounds on the combinations of 
 today's density parameters $(x_1^{(0)}, x_2^{(0)})$ and $(x_3^{(0)}, x_4^{(0)})$.
The colored regions correspond to the parameter space  
constrained by the Planck (red) and PBRS (blue) datasets
at $68\%$ (inside) and $95\%$ (outside) CL limits. 
\label{fig:contours_xi} 
}
\end{figure}

In Table \ref{tab:best_fit_model}, we show the marginalized values of today's four 
density parameters $x_i^{(0)}$ with 95 \% confidence level (CL) limits. 
In Fig.~\ref{fig:contours_xi}, we plot the observationally allowed regions 
derived by two combinations of datasets with the $68\%$ and $95\%$ CL 
boundaries. The best-fit values of $x_1^{(0)}$ and $x_2^{(0)}$ constrained 
by the Planck data are not affected much by including the datasets of 
BAO, SN~Ia, and RSDs.
In the observationally allowed region we have $x_1^{(0)}<0$ and $x_2^{(0)}>0$, 
but there are neither ghosts nor Laplacian instabilities in the constrained 
parameter space (as in the ghost condensate model \cite{Arkani}).

\begin{table}[t!]
\centering
\begin{tabular}{|c|c|c|}
\hline
Parameters  & Planck & PBRS \\
\hline
\hline 
 &  &   \\[-8pt]
$x_1^{(0)}$ & $ -1.32^{+0.21}_{-0.12}~(-1.25) $ & $ -1.35^{+0.01}_{-0.06}~(-1.25) $  \\[-8pt]
 &  &   \\
\hline
 &  &   \\[-8pt]
$x_2^{(0)}$ & $ 1.85^{+0.33}_{-0.69}~(1.62)  $ & $ 1.98^{+0.14}_{-0.29}~(1.68) $  \\[-8pt]
 &  &   \\
\hline
 &  &   \\[-8pt]
$x_3^{(0)}$ & $0.16^{+0.54}_{-0.18}~(0.34) $ & $ 0.07^{+0.2}_{-0.1}~(0.27)  $  \\[-8pt]
 &  &   \\
\hline
 &  &   \\[-8pt]
$x_4^{(0)} (\cdot 10^{-6})$ & $ 0.7^{+2.2}_{-1.8}~(0.15)  $ & $ 0.3^{+0.7}_{-0.6}~(0.54)  $ \\[-8pt]
 &  &   \\
\hline
\end{tabular}
\caption{Marginalized values of the model parameters $x_i^{(0)}$ and 
their 95\,\% CL bounds, derived by Planck and PBRS datasets. 
In parenthesis, we also show the maximum likelihood values 
of these parameters.}
\label{tab:best_fit_model}
\end{table}

\begin{figure*}
\includegraphics[width=.9\textwidth]{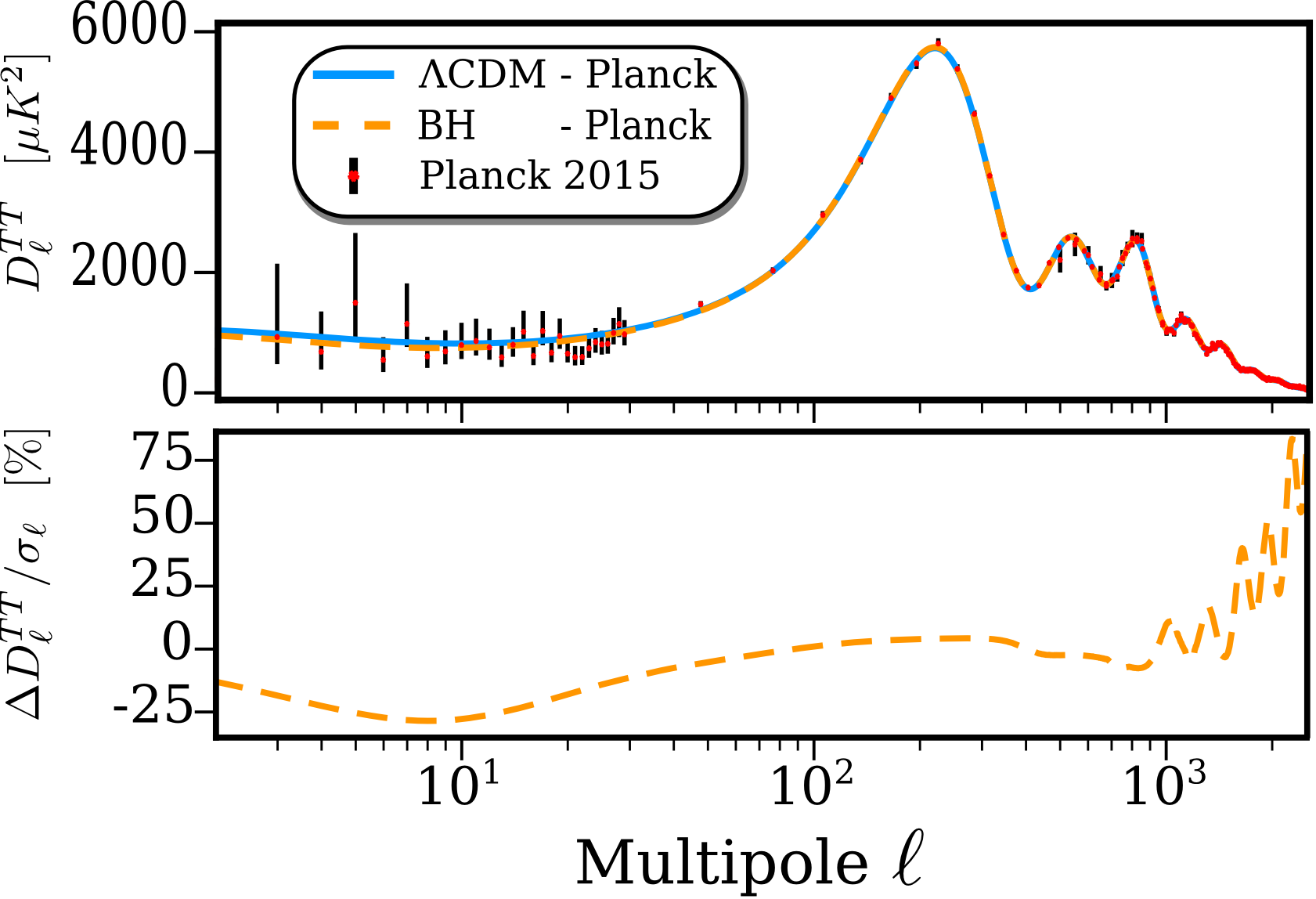}
\caption{(Top) Best-fit CMB TT power spectra 
$D_\ell^{\rm TT} = \ell(\ell+1)C_\ell^{\rm TT}/(2 \pi )$ 
for BH and $\Lambda$CDM, obtained with the Planck dataset. 
The model parameters used for this plot are given in Tables \ref{tab:best_fit_model} 
and \ref{tab:best_fit_cosmo}. 
For comparison, we plot the data points from the Planck 2015 release \cite{Ade:2015xua}. 
(Bottom) Relative difference of the best-fit TT power spectra, in units of the 
cosmic variance $\sigma_\ell = \sqrt{2/(2 \ell+1)} C_\ell^{\Lambda {\rm CDM}}$. 
 See Sec.~\ref{Sec:results} for the difference between the best-fit 
BH and $\Lambda$CDM models.
\label{fig:bestfitTT} 
}
\end{figure*}

\begin{figure}
\includegraphics[width=.49\textwidth]{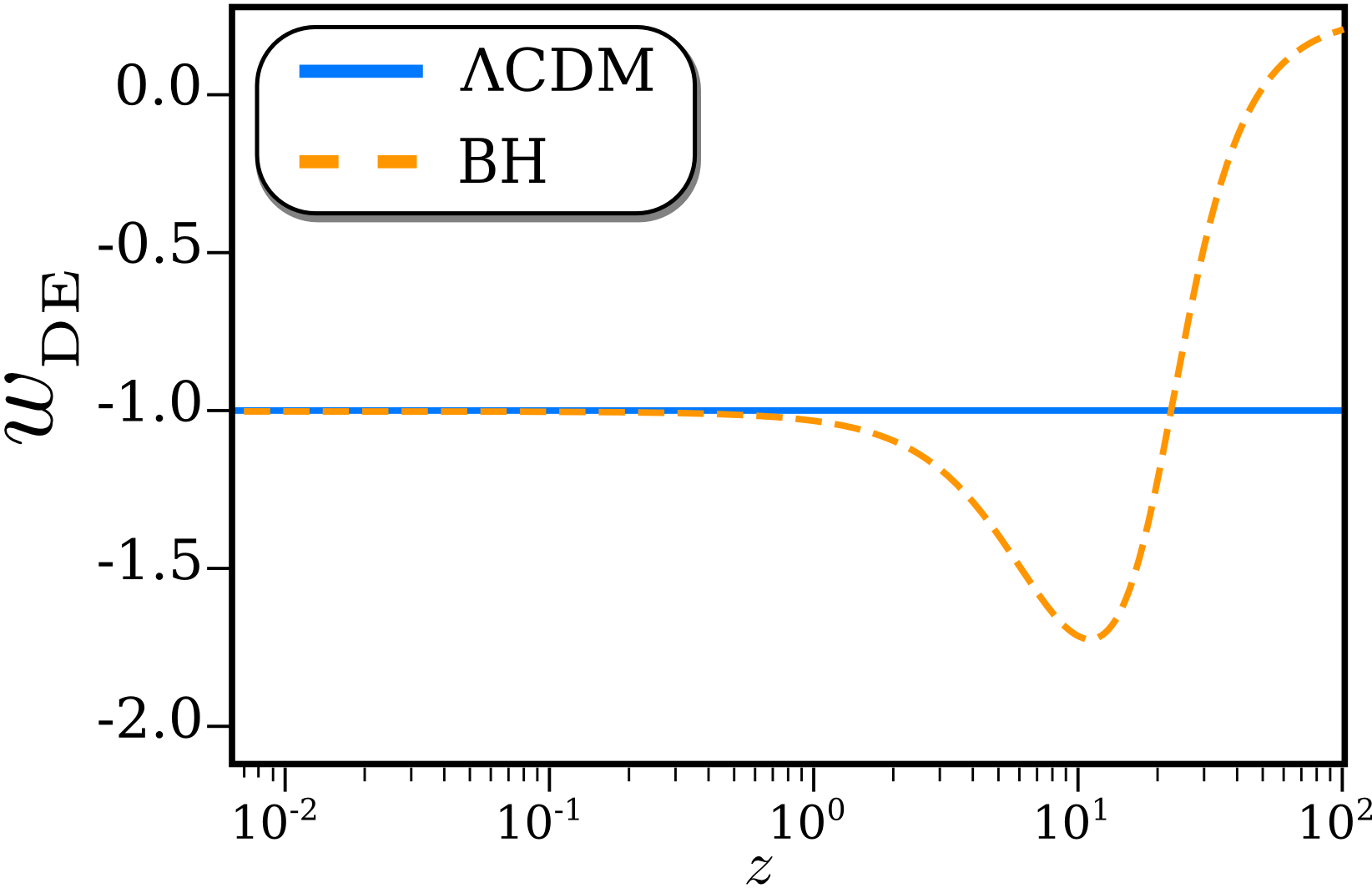}
\caption{Best-fit evolution of the dark energy equation of state 
$w_{\rm DE}$ for BH and $\Lambda$CDM, obtained from the PBRS analysis.
The model parameters used for this plot are given in Tables \ref{tab:best_fit_model} 
and \ref{tab:best_fit_cosmo}.
 In the best-fit BH, $w_{\rm DE}$ first enters the region $w_{\rm DE}<-1$ 
and then it finally approaches the asymptotic value $w_{\rm DE}=-1$.
\label{fig:bestfitwDE} 
}
\end{figure}

With the Planck data alone, the 95 \% CL upper bound on $x_3^{(0)}$ is close to 1, but 
the PBRS datasets give the tighter limit $x_3^{(0)} \le 0.27$ at  95 \% CL. 
The maximum likelihood value of $x_3^{(0)}$ derived with the Planck data is 0.34, 
which is similar to the corresponding value 0.27 constrained with PBRS. 
 The non-vanishing best-fit value of $x_3^{(0)}$ is attributed to
the facts that, relative to $\Lambda$CDM,  
(i) the Galileon term can suppress the large-scale ISW tale, and
(ii) the modified background evolution gives rise to the TT power 
spectrum showing a better fit to the Planck CMB data at high-$\ell$.
In Fig.~\ref{fig:bestfitTT}, these properties can be seen 
in the best-fit TT power spectrum of the BH model.
Increasing $x_3^{(0)}$ further eventually leads to the enhancement of 
the ISW tale in comparison to $\Lambda$CDM. 
 As we see in BH3 of Fig.~\ref{fig:TT_spectra}, the models with large $x_3^{(0)}$ 
do not fit the TT power spectrum well at high-$\ell$ either. 
Such models are disfavored from the CMB data (as in the case of covariant 
Galileons \cite{Renk,Peirone}), so that $x_3^{(0)}$ is bounded from above.
The RSD data at low redshifts can be also consistent with the intermediate values 
of $x_3^{(0)}$ constrained from CMB.

In  Fig.~\ref{fig:bestfitwDE}, we  show the evolution 
of $w_{\rm DE}$ for the best-fit BH model. 
As discussed in Ref.~\cite{Kase:2018iwp}, the existence of $x_2$ besides $x_3$ 
prevents the approach to a tracker solution characterized by $w_{\rm DE}=-2$ 
during the matter-dominated epoch. 
The best-fit background solution first enters the region $-2<w_{\rm DE}<-1$ 
in the matter era and finally approaches a de Sitter attractor characterized by 
$w_{\rm DE}=-1$. Thus, the BH and GGC models with $x_2 \neq 0$ alleviate 
the observational incompatibility problem of tracker solutions 
of covariant Galileons \cite{NDT10}. 
For the best-fit BH model, there is the deviation of $w_{\rm DE}$ 
from $-1$ with the value $w_{\rm DE} \approx -1.1$ at the redshift $1<z<3$, 
so the model is different from $\Lambda$CDM even at the 
background level.

From the PBRS datasets, today's value of $x_4$ is constrained to be 
\be
x_4^{(0)}=0.3^{+0.7}_{-0.6} \times 10^{-6} \quad 
(95~\%~{\rm CL}), 
\label{x4}
\ee
so that $|x_4^{(0)}|$ is at most of order $10^{-6}$. 
With the Planck data alone, the upper bound of $|x_4^{(0)}|$ 
is also of the same order. 
This means that the upper limit of $x_4^{(0)}$ is mostly 
determined by the CMB data. 
As we discussed in Sec.~\ref{Sec:perturbations}, 
the CMB TT power spectrum is sensitive to the dominance of $x_4$ 
over $x_{1,2,3}$ in the early cosmological epoch. 
Then, today's value of $x_4$ is also tightly constrained as 
Eq.~(\ref{x4}), which translates to the bound
\be
|\alpha_{\rm H}^{(0)}| \le {\cal O}(10^{-6})\,.
\ee
Apart from the constraint arising from the GW decay to 
dark energy \cite{Creminelli:2018xsv}, 
the above upper limit on $\alpha_{\rm H}^{(0)}$ is the most 
stringent bound derived from cosmological observations so far.

\begin{figure}
\includegraphics[width=.37\textwidth]{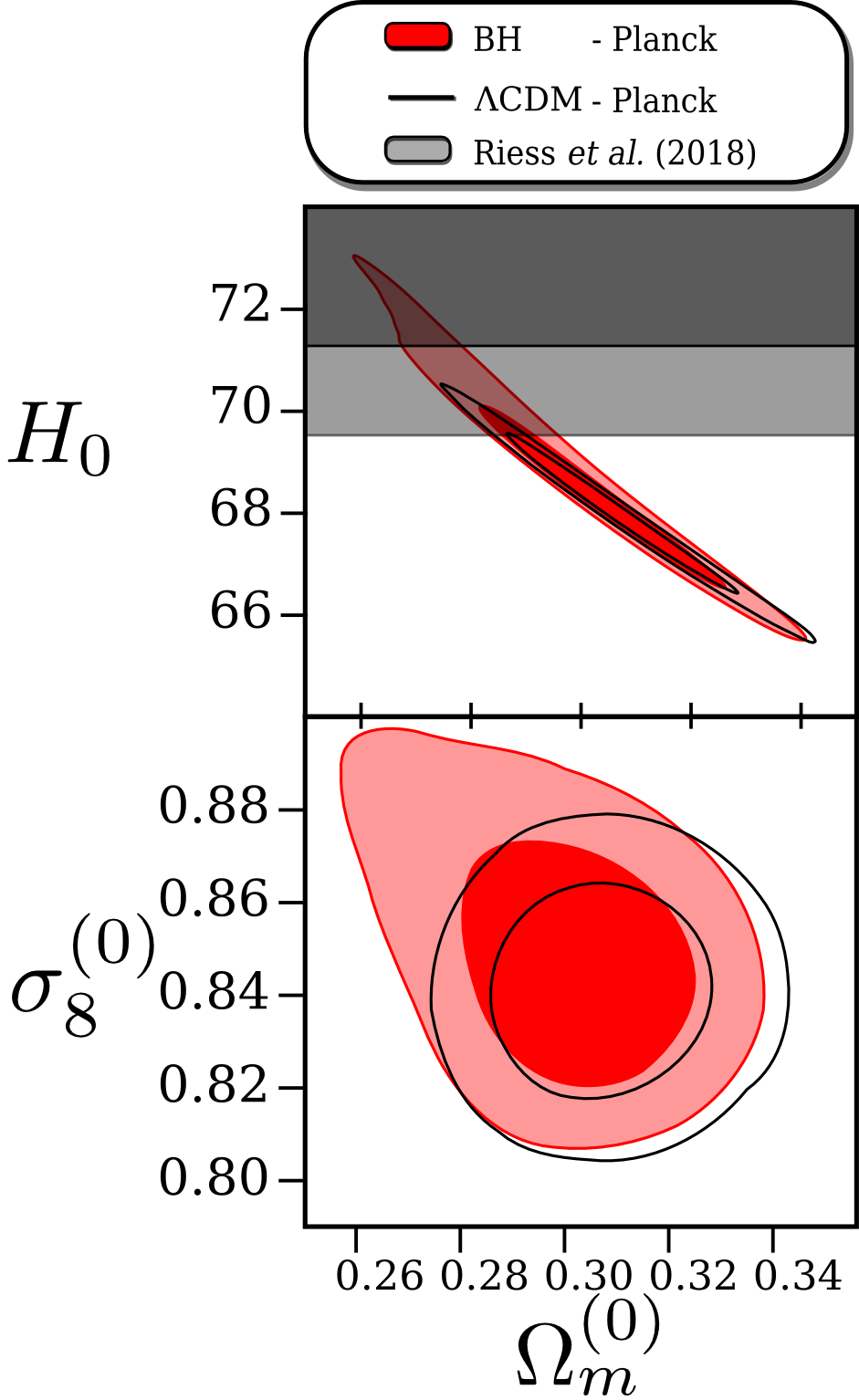}
\caption{The $68\,\%$ and $95\,\%$ CL two-dimensional bounds on 
$(H_0, \Omega_m^{(0)})$ (top) and $(\sigma_8^{(0)}, \Omega_m^{(0)})$ 
(bottom) constrained by the Planck 2015 data, with the unit 
km\,sec$^{-1}$\,Mpc$^{-1}$ for $H_0$.
The observational bounds on BH and $\Lambda$CDM models are shown as 
the red and black colors, respectively. 
In the top panel, the grey bands represent the 
$68\,\%$ and $95\,\%$ CL bounds on $H_0$ 
derived by its direct measurement at low redshifts \cite{Riess:2018uxu}. 
 See the last paragraph of Sec.~\ref{Sec:results} for 
the discussion of likelihood results.
\label{fig:H0s8}}
\end{figure}

In Table \ref{tab:best_fit_cosmo}, we present the values of 
$H_0$, $\sigma_8^{(0)}$, and $\Omega_m^{(0)}$ constrained from 
the datasets of Planck and PBRS for the BH and $\Lambda$CDM models.
The bounds on $H_0$, $\sigma_8^{(0)}$, and $\Omega_m^{(0)}$ derived 
with the PBRS datasets are similar to those in $\Lambda$CDM. 
In Fig.~\ref{fig:H0s8}, we also plot the two-dimensional observational 
contours for these parameters constrained by the Planck data.
The direct measurements of $H_0$ at low redshifts \cite{Riess:2018uxu} give the bound 
$H_0>70$~km\,sec$^{-1}$\,Mpc$^{-1}$, whereas the Planck data tend to 
favor lower values of $H_0$.
Thus, as in the case of $\Lambda$CDM, the BH model does not 
alleviate the tension of $H_0$ between the Planck data and its 
local measurements. The similar property also holds for $\sigma_8^{(0)}$, 
where the Planck data favor higher values of $\sigma_8^{(0)}$ than those constrained in low-redshift measurements. 
We can also put further bounds on $\sigma_8^{(0)}$ by using the datasets of weak lensing measurements, such as 
KiDS~\cite{Hildebrandt:2016iqg,Kuijken:2015vca,deJong:2015wca}. 
For this purpose, we need to take non-linear effects into account in the MCMC analysis, which 
is beyond the scope of the current paper.

\begin{table}[t!]
\centering
\begin{tabular}{|c|c|c|c|}
\hline
Parameter & Model & Planck & PBRS \\
\hline
\hline
 & BH & $ 68.7^{+3.2}_{-2.8}~(69.6)  $ & $ 68.0^{+1.1}_{-1.1}~(68.2) $  \\[-6pt]
$H_0$ &  &   &    \\[-6pt]
& $\Lambda$CDM & $  67.9 \pm 2.0~(67.6)  $ & $ 68\pm 1~(68) $  \\
\hline
& BH & $0.849^{+0.037}_{-0.035}~(0.87) $ & $0.84 \pm 0.03~(0.84)$  \\[-6pt]
$\sigma_8^{(0)}$  &    &   &   \\[-6pt]
&  $\Lambda$CDM & $0.841 \pm 0.03~(0.83)$ & $ 0.84 \pm 0.03~(0.84) $ \\
\hline
&  BH & $0.300^{+0.033}_{-0.034}~(0.28)  $ & $0.306^{+0.014}_{-0.014}$~(0.30)  \\[-6pt]
$\Omega_m^{(0)}$  &   &   &   \\[-6pt]
&  $\Lambda$CDM & $ 0.30 \pm 0.03~(0.31) $ & $ 0.31 \pm 0.01~(0.31) $ \\
\hline
\end{tabular}
\caption{Marginalized values of $H_0$, $\sigma_8^{(0)}$, and $\Omega_m^{(0)}$
and their $95\%$ CL bounds in the BH and $\Lambda$CDM models, 
derived by Planck and PBRS datasets. 
The unit of $H_0$ is km\,sec$^{-1}$\,Mpc$^{-1}$.
In parenthesis, we also show maximum likelihood 
values of these parameters.}
\label{tab:best_fit_cosmo}
\end{table}

\subsection{Model Selection}
\label{Sec:model_selection}

The BH model has three more parameters compared to those in $\Lambda$CDM. 
This means that the former has more freedom to fit the model better with the data.
In order to study whether the former is statistically favored over the latter, 
we compute the Deviance Information Criterion (DIC) \cite{RSSB}:
\be
{\rm DIC} = \chi^2_{\rm eff} (\hat{\theta}) + 2 p_{\rm D}\,,
\label{DIC}
\ee
where $\chi^2_{\rm eff} (\hat{\theta}) = -2 \ln \mathcal{L}(\hat{\theta})$, 
and $\hat \theta$ is a vector associated with model parameters  
maximizing the likelihood function $\mathcal{L}$. 
The quantity $p_{\rm D}$ is defined by 
$p_{\rm D} = \bar{ \chi}^2_{\rm eff} ({\theta}) -  \chi^2_{\rm eff} (\hat{\theta})$, 
where the bar represents an average over the posterior distribution. 
{}From its definition, the DIC accounts for the goodness of fit, 
$ \chi^2_{\rm eff} (\hat{\theta})$, and the Bayesian complexity of the model, 
$p_{\rm D}$. The complex models with more free parameters give
larger $p_{\rm D}$. To compare the BH model with the 
$\Lambda$CDM model, we calculate
\be
\Delta {\rm DIC} =  {\rm DIC} _{\rm BH} - {\rm DIC} _{\Lambda{\rm CDM}}\,.
\ee
If $\Delta {\rm DIC}$ is negative, then BH is favored over $\Lambda$CDM.
For positive $\Delta {\rm DIC} $, the situation is reversed. 

In Table \ref{tab:model_selection}, we present the relative differences of $\Delta \chi_{\rm eff} ^2$ and 
$\Delta {\rm DIC}$ in BH and GGC models, as compared to $\Lambda$CDM. 
Since $\Delta \chi_{\rm eff} ^2$ are always negative, these models provide 
the better fit to the data relative to $\Lambda$CDM.
In particular, we find that $\Delta \chi_{\rm eff} ^2$ constrained by the Planck data 
alone are smaller than those derived with the PBRS datasets.
 This preference of BH over $\Lambda$CDM by the Planck data
arises from combined effects of the suppressed large-scale ISW tale 
caused by the Galileon term and the modified high-$\ell$ TT power spectrum 
induced by the different background evolution 
relative to $\Lambda$CDM (as shown in Fig.~\ref{fig:bestfitTT}).
The former contributes by $\sim 20\,\%$ to a better $\chi^2_{\rm eff}$, 
while the latter to the remaining $\sim 80\,\%$.
We note that a further lowering of the ISW tail is limited by the shift of acoustic peaks 
at high-$\ell$. Such modifications are also subject to further constraints from 
the datasets of BAO and SN~Ia, but the values of $\Delta \chi_{\rm eff} ^2$ constrained 
with the PBRS datasets are still negative in both BH and GGC models.

According to the DIC, the BH model is slightly disfavored over $\Lambda$CDM with 
the PBRS datasets. The GGC model, which has one parameter less than those 
in BH, is favored over $\Lambda$CDM with both Planck and PBRS datasets. 
This implies that the existence of an additional parameter $x_4$ does not 
contribute to provide better fits to the data. Indeed, today's value of $x_4$ is severely 
constrained as Eq.~(\ref{x4}) mostly from the CMB data.  
At the same time, this implies that there are no observational signatures 
for the deviation $\alpha_{\rm H}$ from Horndeski theories. 
It is interesting to note that the GGC model, which belongs to a sub-class 
of Horndeski theories, is statistically favored over $\Lambda$CDM even with 
two additional parameters, but this property does not persist in the BH 
model due to the extra beyond-Horndeski term $\alpha_{\rm H}$ 
modifying the cosmic expansion and growth histories.

\begin{table}
\centering
\begin{tabular}{|c|c|c|c|}
\hline
Model &
Dataset &  $\Delta \chi_{\rm eff} ^2$ & $\Delta {\rm DIC} $\\
\hline
\hline
BH & Planck &  -4.7 & 0.25\\
\hline
BH & PBRS &  -1.8 & 0.1 \\
\hline
GGC & Planck &  -4.8 & -2.5\\
\hline
GGC & PBRS &  -2.8 & -0.6\\
\hline
\end{tabular}
\caption{Model comparisons in terms of $\Delta \chi_{\rm eff} ^2$ and $\Delta {\rm DIC}$. 
As the reference model, we use the value $ \chi_{\rm eff} ^2$ in $\Lambda$CDM. The results for GGC are taken from Ref. \cite{PBFT}.}
\label{tab:model_selection}
\end{table}

\section{Conclusion}\label{Sec:conclusion}

We studied observational constraints on the BH model given by the action (\ref{action}) 
with the functions (\ref{model}). This model belongs to a sub-class of GLPV theories 
with the tensor propagation speed squared $c_t^2$ equivalent to 1.
The deviation from Horndeski theories is weighed by the dimensionless 
parameter $\alpha_{\rm H}=4x_4/(5-x_4)$, where $x_4$ is defined 
in Eq.~(\ref{dimensionless_functions}). 
The BH model also has the $a_2X^2$ and $3a_3X \square \phi$ terms in 
the Lagrangian, which allow the possibility for approaching a de Sitter attractor 
from the region $-2<w_{\rm DE}<-1$ without reaching a tracker solution 
($w_{\rm DE}=-2$).

Compared to the standard $\Lambda$CDM model, the beyond-Horndeski 
term $x_4$ can change the background cosmological dynamics in the early Universe.
Since the Hubble expansion rate $H$ is modified by the non-vanishing 
$x_4$ term, this leads to the shift of acoustic peaks of CMB temperature 
anisotropies at high-$\ell$, see BH1 in Fig.~\ref{fig:TT_spectra}.
Moreover, as we observe in Fig.~\ref{fig:BH1k}, the early-time dominance 
of $x_4$ over $x_{1,2,3}$ leads to the modified evolution of 
gravitational potentials $\Psi$ and $\Phi$ in comparison to $\Lambda$CDM, 
whose effect is more significant for small-scale perturbations. 
This modification also affects the evolution of radiation perturbations 
and the early-time ISW effect.
As a result, the amplitude of CMB acoustic peaks is changed by 
the $x_4$ term. These modifications allow us to put bounds 
on the deviation from Horndeski theories.

The cubic Galileon existing in the BH model leads to the modified growth 
of matter perturbations and gravitational potentials at low redshifts. 
Provided that $x_4$ is subdominant to $x_{1,2,3}$, the dimensional quantities 
$\mu$ and $\Sigma$, which characterize the gravitational interactions with matter 
and light respectively, are given by Eq.~(\ref{musi}) under the quasi-static approximation 
deep inside the sound horizon. 
Thus, the Galileon term $x_3$ enhances the linear growth 
of perturbations without the gravitational slip ($\mu \simeq \Sigma>1$).
This enhancement can be seen in the lensing power spectrum 
$D_{\ell}^{\phi \phi}$ plotted in Fig.~\ref{fig:lensing_spectra}.

For the CMB temperature anisotropies, the late-time modified growth of perturbations 
caused by the cubic Galileon manifests itself in the large-scale ISW tale. 
The ISW effect is attributed to the variation of 
the lensing gravitational potential $\Psi+\Phi$ related to the quantity $\Sigma$.
Unlike the $\Lambda$CDM model in which the time derivative 
$\dot{\Psi}+\dot{\Phi}$ is positive, the Galileon term $x_3$ allows the possibility for realizing 
$\dot{\Psi}+\dot{\Phi}$ closer to 0. In this case, the large-scale TT power spectrum is 
lower than that in $\Lambda$CDM, see GGC and BH2 in 
Fig.~\ref{fig:TT_spectra}. 
Moreover, the modified background evolution at low redshifts 
induced by the Galileon leads to the shift of small-scale 
CMB acoustic peaks toward higher multipoles.
If the contribution of $x_3$ to the total dark energy density is increased 
further, the ISW tale is subject to the significant enhancement 
compared to $\Lambda$CDM, together with the large shift of 
high-$\ell$ CMB acoustic peaks (see BH3 in Fig.~\ref{fig:TT_spectra}). 
These large modifications to the TT power spectrum also arise for 
covariant Galileons without the $x_2$ term, 
whose behavior is disfavored from the CMB data \cite{Renk,Peirone}. 
In the BH model, the existence of $x_2$ besides $x_3$ can give rise to 
the moderately modified TT power spectrum being compatible with the data.

We put observational constraints on free parameters in the BH model by 
running the MCMC simulation with the datasets of CMB, BAO, SN~Ia, and RSDs.
With the Planck CMB data, we showed that today's value of 
$x_4$ is constrained to be smaller than the order $10^{-6}$. 
Inclusion of other datasets does not modify the order of 
upper limit of $x_4^{(0)}$, and hence
$|\alpha_{\rm H}^{(0)}| \le {\cal O}(10^{-6})$.
Apart from the bound arising from the GW decay to dark energy, 
this is the tightest bound on $|\alpha_{\rm H}^{(0)}|$ 
derived so far from cosmological observations.

The other dark energy density parameters $x_1^{(0)}, x_2^{(0)}, x_3^{(0)}$ 
are constrained to be in a similar way to 
those derived in Ref.~\cite{PBFT}. 
The best-fit value of $x_3^{(0)}$ is smaller than 
$|x_1^{(0)}|$ and $x_2^{(0)}$ by one order of magnitude. 
This intermediate value of $x_3^{(0)}$ leads to the CMB 
TT power spectrum with modifications at both large and small scales, 
in such a way that the BH model can be observationally favored over $\Lambda$CDM.
The evolution of matter perturbations at low redshifts is not subject to the large 
modification by this intermediate value of $x_3^{(0)}$ in comparison to $\Lambda$CDM, 
so the BH model is also compatible with the RSD data. The best-fit background 
expansion history corresponds to the case in which $w_{\rm DE}$ finally 
approaches $-1$ from the phantom region $-2<w_{\rm DE}<-1$, whose behavior is 
consistent with the datasets of SN~Ia and BAO.
We also showed that, as in the $\Lambda$CDM model, the tensions in $H_0$ 
and $\sigma_8^{(0)}$ between CMB and low-redshift measurements 
are not alleviated for the datasets used in our analysis.  
Future investigations including non-linear effects and additional probes from 
weak lensing measurements will allow us to shed light on the possibility 
for alleviating such tensions in the BH model.

To make comparison between BH and $\Lambda$CDM models, we computed 
the DIC defined by Eq.~(\ref{DIC}) penalizing complex models with 
more free parameters. In BH, there are three additional parameters than 
those in $\Lambda$CDM.
We found that the effective $\chi_{\rm eff}^2$ statistics in BH 
is smaller than that in $\Lambda$CDM for two combinations of datasets 
(Planck and PBRS). This is mostly due to both the suppressed ISW tail in BH and the shifts 
of high-$\ell$ acoustic peaks of the CMB TT power spectrum. 
These combined effects allow the BH model to fit 
the Planck data better. 
According to the DIC, however, there is a slight preference of 
$\Lambda$CDM over BH with both Planck and PBRS datasets.
The beyond-Horndeski term $x_4$ generally works to prevent 
better fits to the data. The GGC model, which corresponds to $x_4=0$ 
with one parameter less than those in BH, is statistically favored over 
$\Lambda$CDM even with the DIC \cite{PBFT}. 
This means that, at least in the BH model, there is no preference for the 
departure from Horndeski theories in cosmological observations.

We have thus shown that the deviation from Horndeski theories is 
severely constrained by the current observational data, especially from CMB. 
In spite of this restriction, the best-fit BH model gives the effective 
$\chi_{\rm eff}^2$ statistics smaller than that in $\Lambda$CDM. 
Moreover, the GGC model with $\alpha_{\rm H}=0$ leads to the smaller 
DIC relative to $\Lambda$CDM, even with two additional parameters. 
Thus, the BH and GGC models can be compelling and viable candidates 
for dark energy.
Further investigations may be performed in several directions. 
In this work we considered massless neutrinos, but we plan to extend the analysis 
to include massive neutrinos and inquire about any degeneracy which can arise 
between such fluid components and modified gravitational interactions.
Moreover, it is of interest to investigate cross-correlations between the ISW signal and 
galaxy distributions, which can be used to place further constraints on 
BH and GGC models.

\acknowledgments

We thank Matteo Martinelli for support in the numerical implementation. 
We are grateful to N.~Bartolo, A.~De Felice, R.~Kase, 
M.~Liguori, S.~Nakamura, M.~Raveri and A.~Silvestri 
for useful discussions and comments.
SP acknowledge support from the NWO and the Dutch Ministry of Education, Culture and Science (OCW), and also from the D-ITP consortium, a program of the NWO that is funded by the OCW.
GB acknowledges financial sup- port from Fondazione Ing. Aldo Gini.
The research of NF is supported by Funda\c{c}\~{a}o para a  Ci\^{e}ncia e a Tecnologia (FCT) through national funds  (UID/FIS/04434/2013), by FEDER through COMPETE2020  (POCI-01-0145-FEDER-007672) and by FCT project ``DarkRipple -- Spacetime ripples in the dark gravitational Universe" with ref.~number PTDC/FIS-OUT/29048/2017.
NF, SP and GB acknowledge the COST Action  (CANTATA/CA15117), supported by COST (European Cooperation in  Science and Technology).
ST is supported by the
Grant-in-Aid for Scientific Research Fund of the JSPS No.~19K03854 and
MEXT KAKENHI Grant-in-Aid for Scientific Research on Innovative Areas
``Cosmic Acceleration'' (No.\,15H05890).


\end{document}